\documentclass[aps,pra,10pt,showpacs,amsmath,showkeys,twocolumn,floatfix,amssymb, preprintnumbers, nofootinbib, superscriptaddress,longbibliography]{revtex4-1}
\usepackage{graphicx}
\usepackage{comment}
\usepackage[usenames]{color}
\usepackage{bm}
\usepackage{ifpdf}
\usepackage{floatrow}
\usepackage{makecell}
\usepackage{subcaption}

\usepackage[normalem]{ulem}
\usepackage[dvipsnames]{xcolor}
\usepackage[utf8]{inputenc}
\usepackage{hyperref}
\hypersetup{
  pdfnewwindow=true,      
  colorlinks=true,        
  linkcolor=PineGreen,    
  citecolor=PineGreen,    
  filecolor=PineGreen,    
  urlcolor=PineGreen      
}
\newcommand{\be}{\begin{eqnarray}}
\newcommand{\ee}{\end{eqnarray}}

\begin{document}

\title{Characteristic determinant approach to the spectrum of one-dimensional $\mathcal{P}\mathcal{T}$-symmetric systems}
\author{Vladimir~Gasparian}
\email{vgasparyan@csub.edu}
\affiliation{Department of Physics and Engineering,  California State University, Bakersfield, CA 93311, USA}

\author{Peng~Guo}
\email{peng.guo@dsu.edu}

\affiliation{College of Arts and Sciences,  Dakota State University, Madison, SD 57042, USA}
\affiliation{Kavli Institute for Theoretical Physics, University of California, Santa Barbara, CA 93106, USA}

\author{Antonio Pérez-Garrido}
\email{antonio.perez@upct.es}
\affiliation{Departamento de F\'isica Aplicada,  Universidad Polit\'ecnica de Cartagena, E-30202 Murcia, Spain}

\author{Esther~J\'odar}
\email{esther.jferrandez@upct.es}
\affiliation{Departamento de F\'isica Aplicada,  Universidad Polit\'ecnica de Cartagena, E-30202 Murcia, Spain}

\begin{abstract}
We obtain a closed form expression for the energy spectrum of $\mathcal{P}\mathcal{T}$-symmetric superlattice systems with complex potentials of periodic sets of two $\delta$-potentials in the elementary cell. In the presence of periodic gain and loss we analyzed in detail a diatomic crystal model, varying either the scatterer distances or the potential heights. It is shown that at a certain critical value of the imaginary part of the complex amplitude, topological states depending on the lattice size and the configuration of the unit cell can be pushed out from the  given region. This may happened at the $\mathcal{PT}$-symmetry breaking (exceptional) points.
\end{abstract}

\maketitle

\section{Introduction}

It is known  that  when the energy momentum dispersion relation are invariant under the joint action of Parity ($x\rightarrow -x$) and Time-reversal ($i\rightarrow -i$) (see, e.g. Refs. \cite{1a,jons} and references therein), 
the eigenvalues of the Hamiltonian are real (so-called $\mathcal{P}\mathcal{T}$-symmetric system), in spite of the fact that the potential is complex. Such a $\mathcal{PT}$-symmetric complex potentials must satisfy the condition $V^{*}(-x) = V (x)$ and their contact with the environment is highly constrained so that gain from the environment and loss to the environment is exactly balanced.  For   optical systems with $\mathcal{PT}$-symmetric properties, the potential $V(x)$ in the Schrödinger
equation is replaced by the refractive index profile $\epsilon^{*}(-x)=\epsilon(x)$ in the Helmholtz equation. The systems with alternating gain and loss regions demonstrate many novel optical properties (for recent reviews see Refs.~\cite{1a,2a}).  The interplay between gain and index modulation has emerged as a fruitful new research area in photonics. The $\mathcal{PT}$ symmetry, found experimental realizations in the field of photonics in artificial materials with 
spatial distributions of real and complex permittivities, shows the
ability of molding the flow of light (see Refs.\cite{3,4,5}). 
 
Most articles devoted to $\mathcal{PT}$-symmetric systems are related to the study of the problems of eigenvalues and their characteristics \cite{3,4,5,1,2} (for example, the number of eigenvalues, reality, exceptional points for which the eigenstates in the general case are not mutually orthogonal, etc.).  Recently the effort has been extended to scattering problems (see, for example, \cite{6,7,xx,xy,zz1,Guo:2022row} and references therein), and even less to the tunneling time problem \cite{peng1,peng2,Guo:2024bar} and Faraday/Kerr effect \cite{Gasparian:2022acf,Guo1}.

The aim of this paper is to carry out a generalization of the method of characteristic determinant, developed for the real and complex potentials in Refs.\cite{GA88,Guo1}, to the spectrum of $\mathcal{P}\mathcal{T}$-symmetric superlattice systems with complex potentials of periodic sets of $\delta$-potentials with $n$ centers in the elementary cell (see Eq.(\ref{pot2a}) below).

The characteristic determinant method is compatible with the transfer matrix method and allows in some cases to carry out efficient analytical calculations when the number of scattering potentials become very large, in contrast to the transfer matrix method, which in such cases becomes practically useless.  It based on conventional scattering formalism, depends on the amplitudes of reflection of a single scatterer only and does not require further tight-binding approximations. 
The key idea is to start from a single cell and gradually build up to a many-cell system by adding one cell at a time mapping the problem of the calculation of the determinant $ D_N $ of order $ N \times N $ ($N$ is a number of individual scattering sites or atoms in a chain). Note that the calculation of the local and average density of states for the sample, the reciprocal of the transmission amplitude, and the characteristic tunneling time of the barrier \cite{Pollak,gas1}, are directly related to the determinant $D_N$.

The model of $\mathcal{P}\mathcal{T}$-symmetric systems of delta potentials, presented  by Eq. (\ref{pot2a}), leads to an exact analytical expression for the energy spectrum of electrons in the infinite square well with  arbitrary $N$ Dirac delta function potentials depending on the values of the gain-loss parameters. 

Particularly for a diatomic crystal, ($n=2$ atoms in a unit cell), varying either the scatterer distances or the potential heights, we obtained a closed form expression for the energy spectrum in terms of the number of cells, the real and imaginary parts of the complex amplitudes of $\delta$-potentials and separation of the potentials.  It is shown, that at a certain critical value of the imaginary part of the complex amplitude the topological states may move from given regions. This may happened at the $\mathcal{PT}$-symmetry breaking (exceptional) points.  Recently, the properties of the topologically protected edge states for the $\mathcal{PT}$ -symmetric extension of the Su-Schrieffer-Heeger model governed by the non-Hermitian Hamiltonian, was studied both numerically and analytically in Ref. \cite{hay}. A topologically nontrivial non-Hermitian trimerized optical lattice is investigated in Ref.~\cite{chin}.

The work is organized as follows. In Section II we formulated the problem of determining the energy spectrum through the characteristic determinant. In Sec. III we consider the technically simple case of the periodic structure with a two delta potentials in a unit cell. For this finite bipartite Kroning-Penney $\mathcal{P}\mathcal{T}$-symmetric models appropriate analytical expressions
for scattering matrix elements and for gain/loss parameters are derived. The main conclusions are summarized in Sec. IV.

\section{Energy spectrum in terms of the characteristic determinant: General statement}

\subsection{Short Summary of the characteristic determinant method for open systems}
It is known that the poles of Green's functions in open systems correspond to the spectrum of excitations, meanwhile in the case of a closed system, the poles coincide with energy spectrum of the system. However, calculating the Green's function (GF) for any system with an arbitrary potential $V(x)$ is a difficult task. Technically, the problem is that the calculation of the GF, which can always be written in the form of a bilinear expansion in eigenfunctions, requires knowledge of the latter, that is, an exact solution of the Schr{\"o}dinger equation.  That's why explicit expressions for GF known only for a few specific potentials (see, for example, Refs. \cite{zatik,eco} and references therein). Surprisingly, it often turns out that calculating the poles (or zeros of the characteristic determinant) of the GF is, from a technical point of view, relatively simpler than calculating the GF itself. For the Kronig-Penney (KP) and tight binding (TB) models, the poles have been calculated in quasi-one and two-dimensional disordered systems without any limitation on the number of impurities and modes in terms of the so-called characteristic determinants in Ref.\ \cite{GA88,Guo1}. This nonperturbative approach, sufficiently
  describe electron (photon) behavior in a random
potential, and allow one to study the energy spectrum and scattering matrix
elements in any system without actually determining the
electron (photon) eigenfunctions. Once characteristic determinant, $D_N$, is obtained,  the determination of the bound states is easy. It can be shown \cite{GA88,Guo1,zatik,zatik1} that the poles of the GF of the whole system (which, as we know, correspond to the bound states) are just the zeroes of the
characteristic determinant. Therefore, to find the bound states of the potential for both models we just will have to solve the equation: $D_N = 0$. The determinant $D_N$ is in general a complex function of the energy $E$. Hence, we need to find
simultaneous zeroes in its real and imaginary parts.

In what follows, we will present some details of characteristic determinant method. 
Based on the Refs.~\cite{GA88,Guo1}, the characteristic determinant $D_N$ of a one-dimensional chain of $N$ $\delta$-potential with complex amplitudes $Z_l\equiv \eta_{1l} +i\eta_{2l}$ and corresponding coordinates $x_l$ is  defined by
\begin{equation}
D_N=\det \left ( M^{(N)}_{n,l} \right ) \;,
\end{equation}
where
\begin{equation}
M_{n,l}^{(N)} =\delta_{n,l}+\frac{iZ_l}{2k} e^{ik|x_l-x_n| },
\qquad 1 \leq ( n,l) \leq N\;. \label{eq:det}
\end{equation}
The $k $ denotes the momentum of electron.   The total energy of an electron $E$ is related to momentum $k$ by $E=k^2$, where the mass of electron is assumed  as $1/2$ in this work.  The characteristic determinant, $D_N$, can also be written as the determinant of a tridiagonal Toeplitz matrix and satisfies the following recurrence relationship:
\begin{equation}
D_N={A_N} D_{N-1} - {B_N} D_{N-2}\;,\label{eq:d0}
\end{equation}
where $D_{N-1}$ ($D_{N-2}$) is the determinant Eq.(\ref{eq:det}) with the $N$-th (and also the $(N-1)$-th) row and column omitted. The coefficients $A_N$, $B_N$ can be obtained from the explicit form of $M_{n,l}^{(N)}$. For $N>1$ we have
\begin{equation}
{A_N}=1+ {B_N}+\frac{iZ_N}{ 2k}
\left(1-e^{2ik(x_{N}-x_{N-1})}\right)\;,
\end{equation}
and
\begin{equation}
{B_N}= \frac{iZ_N}{ Z_{N-1} } e^{2ik(x_{N}-x_{N-1})}\;.
\end{equation}
The initial conditions for the recurrence relations are
\begin{equation}
D_0=1,\quad D_{-1}=0,\quad {A_1}=1+ \frac{iZ_1}{ 2k} ,\quad  B_1 =0\;.
\end{equation}

The transmission amplitude, $t$, is the inverse of the characteristic determinant $D_N$ multiplied by the phase accumulated during the transmission, i.e.,
\begin{equation}
t={e^{ik(x_N-x_1)}D_N^{-1}}\;. \label{t}
\end{equation}
While the left and right propagating reflection amplitudes $r_L$ and $r_R$ are given by
\begin{equation}
r_L=- \frac{2k}{ iZ_{1}} \frac{D_N-D_{-1+N}}{ D_{N}}-1 \equiv
i \frac{\partial \ln t  }{ \partial \frac{Z_1}{2k}}-1, \label{rL}
\end{equation}
and
\begin{equation}
r_R=- \frac{2k }{ iZ_{N}}  \frac{D_N-D_{N-1} }{ D_{N}}-1 \equiv
i \frac{\partial \ln t  }{ \partial \frac{Z_N}{2k}}-1, \label{rR}
\end{equation}
respectively. Here $D_{-1+N}$ is the characteristic determinant without the first delta function (i.e. $Z_1=0$) and $D_{N-1}$ is the characteristic determinant without the last delta function (i.e. $Z_N=0$).

\subsection{Quantization condition of closed systems with hard wall boundary conditions}
In order to further investigate the energy spectrum of the closed system with hard wall boundary conditions, let us, following\cite{GAO}, write explicitly the dependence of the characteristic determinant $D_N$ on $Z_1$ and $Z_N$ for our general system of multiple $\delta$ potentials. From the recurrence relations for the characteristic determinant, Eq.(\ref{eq:d0}), applied to both ends, we can
rewrite $D_N$ in the following way ($m\equiv N-2$):
\begin{align}
&D_N(Z_1,Z_N) \nonumber \\
& =D_{m}\left (1+\frac{A_m}{2k} iZ_1 +\frac{B_m}{2k} iZ_N +\frac{C_m}{4k^2} Z_1 Z_N \right),
\label{dn}
\end{align}
where $D_{m}$ is the characteristic determinant for the previous potential without the first and the last delta function (i.e., $Z_1=Z_N=0$), and $A_m$, $B_m$ and $C_m$ are coefficients independent of $Z_1$ and $Z_N$ and involving $D_{m}$.
These coefficients are defined as
\begin{align}
A_m &= 1-i\sqrt{1-T_m}e^{i\Theta_1}, \label{A}  \\
B_m &= 1-i\sqrt{1-T_m}e^{i\Theta_2} , \label{B} 
\end{align}
and
\begin{align}
& C_m =2ie^{i \frac{ \Theta_1+\Theta_2 }{ 2 } } \nonumber \\
& \times \left[\sin \left ( \frac{\Theta_1+\Theta_2}{2} \right )
+\sqrt{1-T_m}\cos \left( \frac{\Theta_1-\Theta_2 }{2} \right )\right],
\label{C}
\end{align}
where $T_m=t_mt^*_m$ is the transmission coefficient of the system of $m$
contact potentials. The transmission amplitude through the system, $t_m$,  is determined by the Eq. (\ref{t}) with $D_N$ replaced by $D_m$, and $(x_N-x_1)$ replaced by $(x_{N-1}-x_2)$. The phases appearing in the previous equations are defined as
\begin{align}
\Theta_1 &=\varphi_m+\varphi_{a,m}+2k(x_2-x_1) , \nonumber \\
\Theta_2 & =\varphi_m-\varphi_{a,m}+2k(x_N-x_{N-1}),
\end{align}
where $\varphi_m$ is the phase accumulated in a transmission event and $\varphi_{a, m}$ is the phase characterizing the asymmetry between the reflection to the left
and to the right of the block with $m$ contact potentials. 

Note that the expression in Eq.(\ref{dn}) for $D_N(Z_1,Z_N)$ is written for an open system. To close the system with hard wall boundary conditions on both ends of chain and obtain the desired expression for the energy spectrum, it is necessary to increase the amplitudes of the delta potentials $Z_1$ and $Z_N$. In the limit of $Z_1$ and $Z_N$ tending to infinity, we keep only the leading term $C_m\frac{Z_ 1 Z_ N}{4k^2}$ in $D_N(Z_1,Z_N)$. The zeros of the $C_m$   yields the quantization condition that determines the energy spectrum of systems confined by hard wall boundary conditions, hence a compact expression of quantization condition can be obtained 
\begin{align}
& \sin\left (\varphi_{m}+k(x_{N}-x_{N-1}-x_1+x_2) \right) \nonumber \\
&+
\sqrt{1-T_m}\cos\left(\varphi_{a,m}-k(x_{N}-x_{N-1}-x_2+x_1)\right)=0, \label{eq.1}
\end{align}
where $x_{N}$ and $x_1$ are the ends of the chain where the hard walls are located, and $x_{2}$ and $x_{N-1}$ are the first and last coordinates of the block with  $m$ contact delta potentials.

We remark that  though we arrived at the above equation using the $\delta$-potential model,  it can be generalized to an arbitrarily shaped potential by following Ref. \cite{aronov}. Therefore, Eq. (\ref{eq.1}) is a fairly general expression regardless of the shape of the potential under consideration. It can be used to find the energy spectrum of electrons in any finite closed 1D system divided into three blocks, see Section \ref{sec:KPmodel}.

\section{Finite bipartite Kroning-Penney $\mathcal{P}\mathcal{T}$-symmetric model}\label{sec:KPmodel}

In what follows we analyze in detail the spectrum of one-dimensional (1D) finite
Kronig-Penney complex non-conservative model, by using quantization condition given in Eq.(\ref{eq.1}).

\begin{figure}
\includegraphics[width=.99\textwidth]{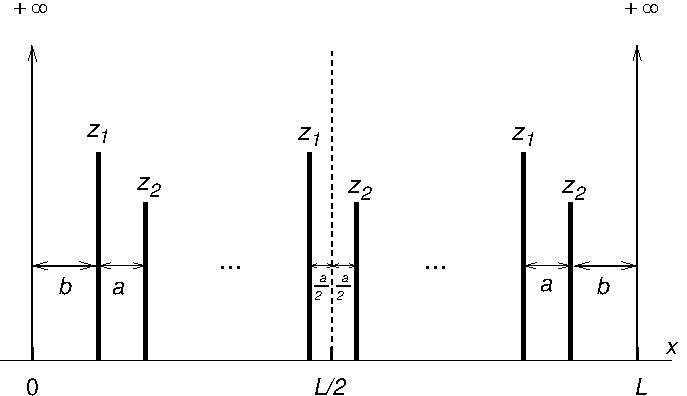}
\caption{Demo plot of a $\mathcal{PT}$-symmetric system with two $\delta$ potentials in a unite cell, described by Eq.(\ref{pot2a}).}
\label{demoplot}
\end{figure}

The mathematical model of the $\mathcal{P}\mathcal{T}$-symmetric superlattice systems with complex potentials
discussed below is the infinite well with $2 M$ Dirac delta function potentials with complex strengths that are distributed in the following way:
\begin{equation}
V(x) =\sum^{M}_{n=1} \bigg [   Z_1\delta\left (x - (  n d - a) \right )   +    Z_2\delta\left(x-  n d \right )  \bigg ], \label{pot2a}
\end{equation}
where $d$ is the lattice period, $a$ is the distance between two successive potentials $Z_1$ and $Z_2$, $Z_1=Z^*_{2}\equiv \eta_1+i\eta_2$ ($\eta_2$ represents non-Hermitian degree). The system has $M$ cells and $ 2M$ scatterers, respectively. The hard walls are placed at: $x  = 0$ and $x  =L$, where $L=b + M d$ and  $b= d -a$, see Fig.~\ref{demoplot},   and   the wavefunction vanishes when $x <0$ and $x > L$. A closed system  with $2M$ finite delta potentials located in-between two infinite potentials is obtained.  The $\mathcal{PT}$ symmetry requirement is equivalent to: $$V \left (x-\frac{L}{2} \right )=V^* \left ( -x+ \frac{L}{2} \right ).$$

The transmission coefficient $T_{m}$ through the system, where $m = 2M $. The phase $\varphi_{m}$, accumulated during the transmission and the phase difference $\varphi_{a, m}$ between the reflection from the left and from the right, entering in the Eq. (\ref{eq.1}), can be evaluated  by using the method of the
characteristic determinant, developed  in Refs.~\cite{GA88,Guo1}. 

Before going into the details of the 
quantization condition given by equation (\ref{eq.1}), we present several useful relations characterizing the 
$\mathcal{P}\mathcal{T}$-symmetric model.

After some algebraic
manipulations, we find   
\begin{equation}
T^{-1}_{m}\equiv |D_{m}|^2=1+\sqrt{|D_2|^2-1}\frac{\sin^2 (M\beta d) }{\sin^2 (\beta d) },\label{T2m} 
\end{equation} 
\begin{equation}
\varphi_{m} =-kb -\arctan \left [ \frac{ \mbox{Im}\left[e^{-ikd}D_2\right]}{\mbox{Re}\left[e^{-ikd}D_2\right]} \right ],\label{spect2a0}
\end{equation}
and
\begin{equation}
\varphi_{a,m} =2\arctan \left [ \frac{\left(\frac{Z_2}{2k}-\frac{Z_1}{2k}\right)\sin(ka)}{\left(\frac{Z_2}{2k}+\frac{Z_1}{2k}\right)\cos (ka)+2\frac{Z_1}{2k}\frac{Z_2}{2k}\sin(ka)} \right ],\label{spect2a1}
\end{equation}
where again $b= d -a$ is the distance between the left (right) hard wall and the first (last) delta potential: $x_{2M}-x_{2M-1}$=$x_2-x_1\equiv b$. The characteristic determinant $D_2$ of one cell with two delta potentials is defined by  
\begin{equation}
 D_2=\det\left(
\begin{array}{ll}
1+i\frac{Z_1}{2k}&i\frac{Z_2}{2k}e^{ika}\\
i\frac{Z_1}{2k}e^{ika}&1+i\frac{Z_2}{2k}\\
\end{array} \right). \label{d2}
\end{equation}

Taking into account the periodicity of the system and using recurrence relations in Eq.(\ref{eq:det}), one can also show that the important quantity $D_{m}$ for a diatomic crystal may be written in the form \cite{est}
\begin{equation}
D_{m}=e^{ikMd}\left\{\cos(M\beta d)+iIm\left[e^{-ikd}D_2\right]\frac{\sin(M\beta d)}{\sin(\beta d)}\right\}, \label{dm21} 
\end{equation}
where $\beta$ plays the role of quasimomentum for the diatomic crystal and is given by the equation
\begin{align}
& \cos(\beta d)\equiv \mbox{Re}\left[e^{-ikd}D_2\right] =\cos(kd) \nonumber \\
&+\left(\frac{Z_1}{2k} +\frac{Z_2}{2k}\right)\sin(kd)+
2\frac{Z_1}{2k}\frac{Z_2}{2k}\sin(ka)\sin{k(d-a)}. \label{spect2xa}
\end{align}
As for the term $\mbox{Im}\left[e^{-ikd}D_2\right]$ in Eq. (\ref{dm21}), it can be written in the form 
\begin{align}
&  \mbox{Im} \left[e^{-ikd}D_2\right]  \equiv \left(\frac{Z_1}{2k}+\frac{Z_2}{2k}\right)\cos(kd)  \nonumber \\
& -\sin(kd)+
2\frac{Z_1}{2k}\frac{Z_2}{2k}\sin(ka)\cos{k(d-a)} .
\end{align}
Finally,  after some algebraic
manipulation,  the Eq.(\ref{eq.1}) yields  a  compact form of  quantization condition:
\begin{equation}
\frac{\sin(kb)\sin((M+1)\beta d)+{\sin(M\beta d)}\sin(ka) }{\sin{\beta d}} =0, \label{C12}
\end{equation}
Eq.(\ref{C12}) is our main general result for the energy spectrum of $\mathcal{P}\mathcal{T}$-symmetric superlattice systems. It holds for both real and complex potentials, extending several well-known results in the literature and help to get even more insight into the mathematical structures of the eigenvalues. 
To demonstrate this explicitly, it is useful to consider the relatively simple finite bipartite Kroning-Penney model discussed, for example, in Refs. \cite{4,benj,resh,many} for real potentials. We want emphasize here that one can use Eq.(\ref{C12}) and Eq.(\ref{spectaa}) (see below) as a starting point in reproducing the most of the main results of Refs. \cite{4,benj,resh,many}, calculated numerically.

\subsection{Real potentials }
In case of simple Kroning-Penney model, that is $Z_1 =Z_2\equiv V_0>0$ (single positive delta potential in a cell: $d=2a$ and $b=a$) the above expression (\ref{C12}) reduced to
\begin{equation}
\sin (ka) \frac{\sin{(m+1)\beta a}}{\sin(\beta a)}=0, \label{spectaa}
\end{equation}
where now $m$ is the number of delta potentials (even or odd) and $\cos( \beta a)$ defined as 
\begin{equation}
\cos(\beta a )=\cos(ka)+\frac{V_0}{2k}\sin(ka) .\label{spect1}
\end{equation}
Note that for particular values of $m=1$ and $m=2$, the above formula reduces to the results found in Refs. \cite{,iran,many}, 
 using a different approach. Note also that the expression in Eq.(\ref{spectaa}) is similar in spirit, but differs in the details of the physical context from some other known results where interference plays a critical role. The class of phenomena to which this applies is the light diffraction through multiple slits, Landauer's resistance between an ideal conductors and a periodic structure, traversal, dwell, and tunneling times of particles (photons or electrons) traveling through periodic systems, etc. For example, the zeros of the energy spectrum of the simple Kroning-Penney model with $m$ delta potentials (Eq.(\ref{spectaa})) formally coincide (in units $e^2/h$) with the zeros of the two terminal Landauer's  resistance between an ideal conductors and a periodic structure with (m+1) delta potentials in Ref.~\cite{GA88}:
\begin{align}
 \rho_{m+1} &=|D_{m+1}|^2-1 \equiv \frac{1-T_{m+1}}{T_{m+1}}\nonumber \\
&=\left(\frac{V_0}{2k}\right)^2\frac{\sin^2 (m+1)\beta a }{\sin^2(\beta a)} . 
\end{align}

\begin{figure*}
 \begin{subfigure}[b]{0.99\textwidth}
\includegraphics[width=0.495\textwidth]{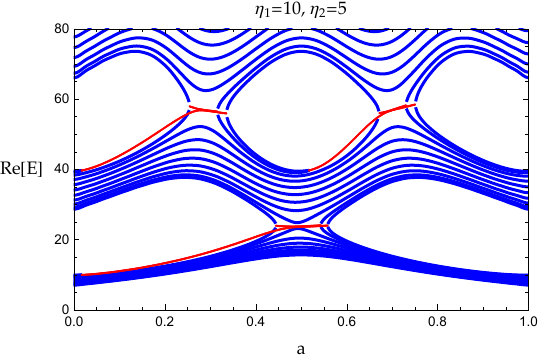}
\includegraphics[width=0.495\textwidth]{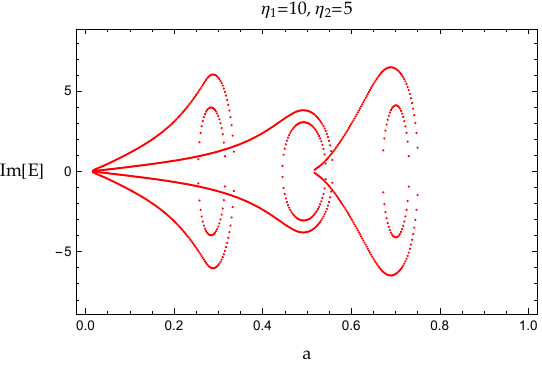}
\caption{Real (left) and Imaginary (right) parts of eigen-energy with $\eta_1=10,\eta_2=5$.} \label{FIG2a}
 \end{subfigure}
 \begin{subfigure}[b]{0.99\textwidth}
\includegraphics[width=0.495\textwidth]{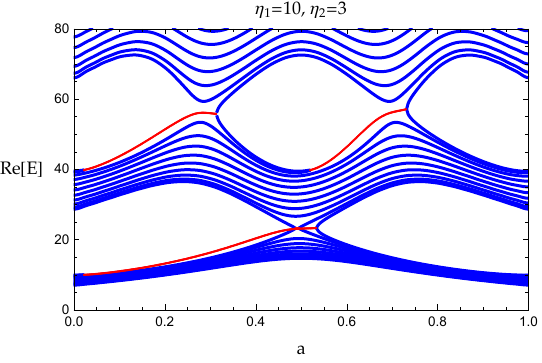}
\includegraphics[width=0.495\textwidth]{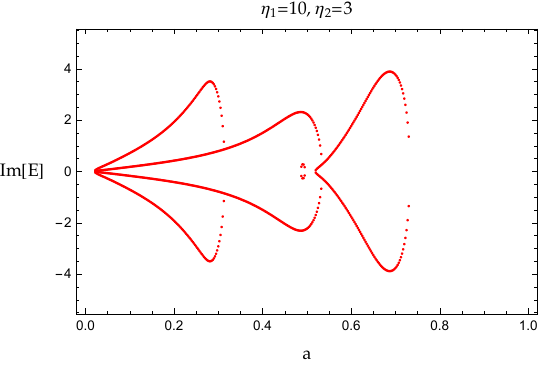}
\caption{Real (left) and Imaginary (right) parts of eigen-energy with $\eta_1=10,\eta_2=3$.} \label{FIG2b}
 \end{subfigure}
 \begin{subfigure}[b]{0.99\textwidth}
\includegraphics[width=0.495\textwidth]{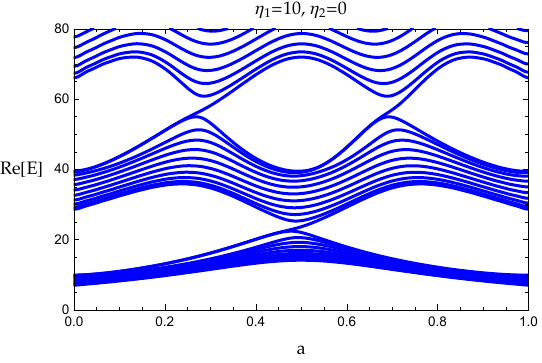}
\caption{Real    eigen-energy with $\eta_1=10,\eta_2=0$.} \label{FIG2c}
 \end{subfigure}
\caption{The real and imaginary part of energy spectrum  as a function of the distance $a=1-b$ between two $\delta$-potentials $z_1$ and $z^*_1$ for $\eta_1=+10$ positive strength and varying $\eta_2$. The real energy solutions are plotted in blue, and the complex energy solutions are plotted in red. The number of cells is $M=10$. }
\label{tau1negativo1}
\end{figure*}

\begin{figure*}
  \begin{subfigure}[b]{0.99\textwidth}
\includegraphics[width=0.495\textwidth]{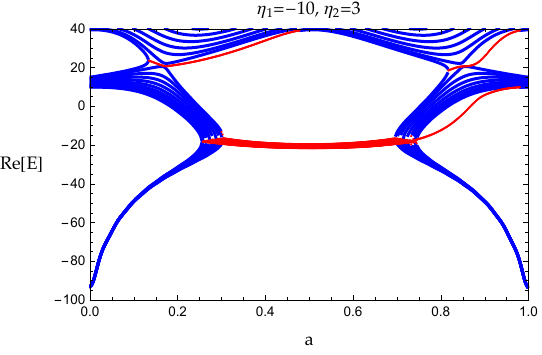}
\includegraphics[width=0.495\textwidth]{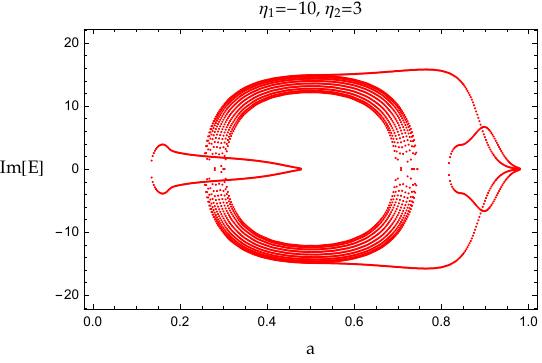}
\caption{Real (left) and Imaginary (right) parts of eigen-energy with $\eta_1=-10,\eta_2=3$.} \label{FIG3a}
 \end{subfigure}
 \begin{subfigure}[b]{0.99\textwidth}
\includegraphics[width=0.495\textwidth]{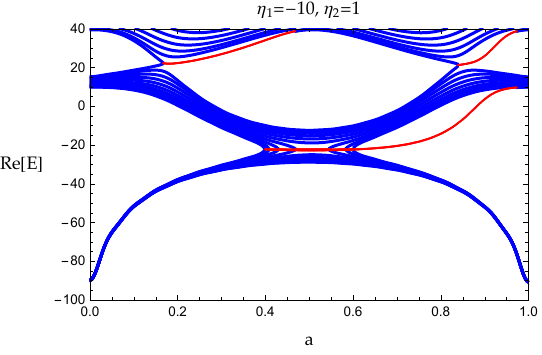}
\includegraphics[width=0.495\textwidth]{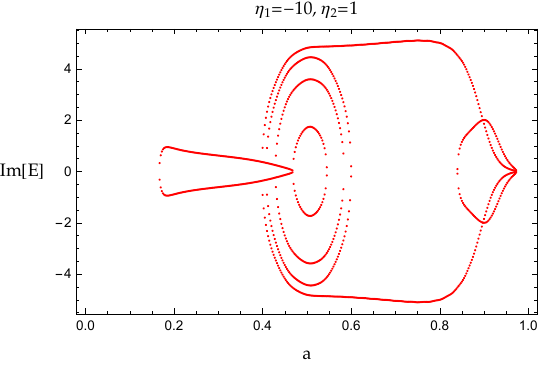}
\caption{Real (left) and Imaginary (right) parts of eigen-energy with $\eta_1=-10,\eta_2=1$.} \label{FIG3b}
 \end{subfigure}
 \begin{subfigure}[b]{0.99\textwidth}
\includegraphics[width=0.495\textwidth]{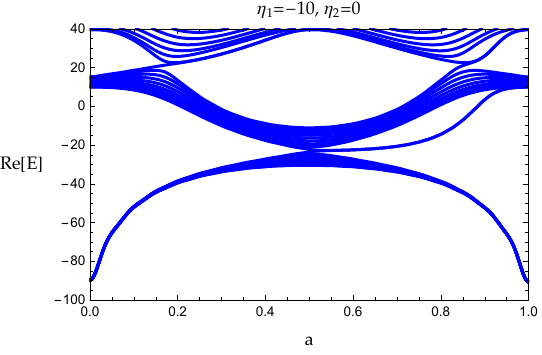}
\caption{Real   eigen-energy with $\eta_1=-10,\eta_2=0$.} \label{FIG3c}
 \end{subfigure}
\caption{The real and imaginary part of energy spectrum as a function of the distance $a=1-b$ between two $\delta$-potentials $z_1$ and $z^*_1$ for $\eta_1=-10$ negative strength and varying $\eta_2$. The real energy solutions are plotted in blue, and the complex energy solutions are plotted in red. The number of cells is $M=10$. }
\label{tau1negativo2}
\end{figure*}

\subsection{Complex $\mathcal{P}\mathcal{T}$-symmetric potentials}

After a brief discussion of the energy spectrum, given by Eq.(\ref{C12}), at real potentials, let us analyze the spectrum when $Z_1=Z^{*}_2 =\eta _1+i\eta_2$ is an arbitrary complex number.

Similar to the above derivation of Eq.(\ref{C12}), one can show that the energy spectrum in case of an arbitrary complex number of $Z$ formally is still given by the same equation. As for $\cos(\beta d)$, we write it explicitly in terms of $\eta_1$ and $\eta_2$:
\begin{equation}
\cos (\beta d)=\cos(kd)+\frac{\eta_1}{k}\sin(kd)+
\frac{\eta^2_1+\eta^2_2}{2k^2}\sin(ka)\sin(kb).\label{spect2x}
\end{equation}
We now analyze  the energy spectrum expression (\ref{C12}) for different $\eta_1$ and $\eta_2$, based on relation in Eq.(\ref{spect2x}), to see a dynamical evolution of the band structure depending on the imaginary portion $\eta_2$, see e.g. Fig.~\ref{tau1negativo1} and Fig.~\ref{tau1negativo2}.  Fig.~\ref{FIG2c} and Fig.~\ref{FIG3c} show the band structure for a chain consisting of $M=10$ cells or $20$ positive $\delta$-like potentials as a function of $a$  with fixed $d =1$ and $\eta_2=0$. The parameter $a$ can be considered as a shift relative to the wall of the box and ultimately play the role of some additional virtual second dimension in studying the topology of the bands, see e.g. Ref.~\cite{kp1}. A similar situation with an additional degree of freedom, that is typical on the topic of topological insulator in one-dimensional system, arises in the case of a tight binding model with a modulated tunneling parameter \cite{report}.  As seen from Fig.~\ref{FIG2c} and Fig.~\ref{FIG3c}, there are states that are separating from the upper band and move to lower band through the forbidden gap.   For  example, the extra level  on the right-hand side in Fig.~\ref{FIG3c} between the first and the second bands is  the signature of topological edge states, see e.g. the discussion of a Bipartite Kronig-Penney model in Ref.\cite{benj}, which would be absent if the hard wall boundary is replaced by an infinite long periodic  system. These extra topological edge levels are also demonstrated in Fig.~\ref{FIG3a} with hard wall boundary condition compared with Fig.~\ref{fig5} for a periodic infinite long system where topological edge states are absent.  In principle, the   Zak phase and the winding number of the reflection coeﬃcient  could be evaluated for topological states, see e.g. Ref.\cite{benj}. The surface, interface or contact states that arise at the boundary of two systems in common gap, can be exponentially localized to the edges. They are determined by the poles of the reflection amplitudes left, $r_L$ in Eq.(\ref{rL}), or right, $r_R$ in Eq.(\ref{rR}). Positions of these levels depend significantly on the matching conditions (or parameter $a$), and in a real crystal, due to the surface roughness, the states must be distributed over the gap, see e.g. Ref.~\cite{tamm}. 

Now let's first look at what happens to the band structure when $\eta_2\ne 0$.

\begin{figure*}
\includegraphics[width=0.48\textwidth]{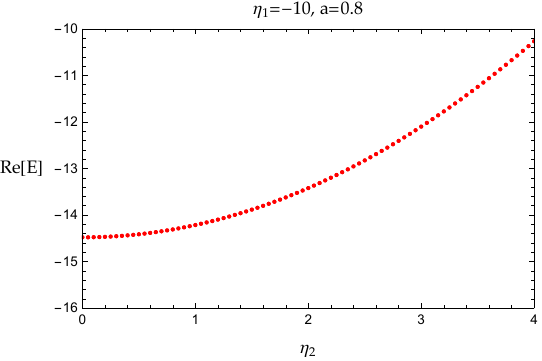}
\includegraphics[width=0.48\textwidth]{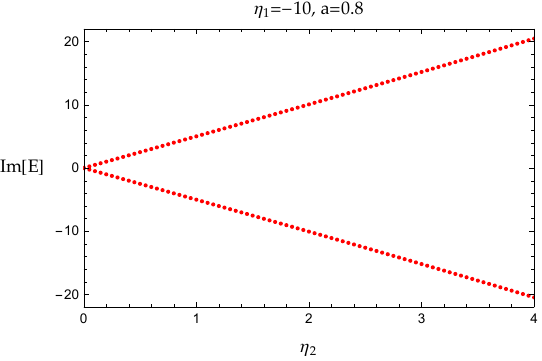}
\caption{The real and imaginary part of the lowest edge state energy at $a=0.8$ in Fig.\ref{tau1negativo2} as the function of $\eta_2 \in [0,4]$.}
\label{fig4}
\end{figure*}

\begin{figure*}
\includegraphics[width=0.48\textwidth]{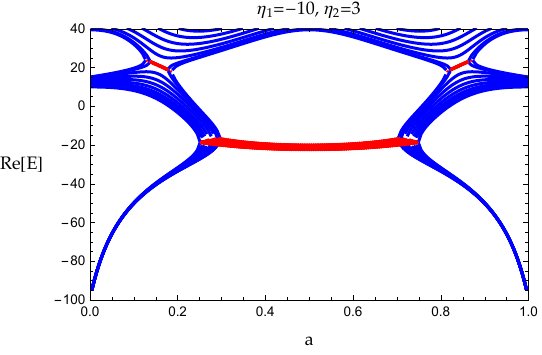}
\includegraphics[width=0.48\textwidth]{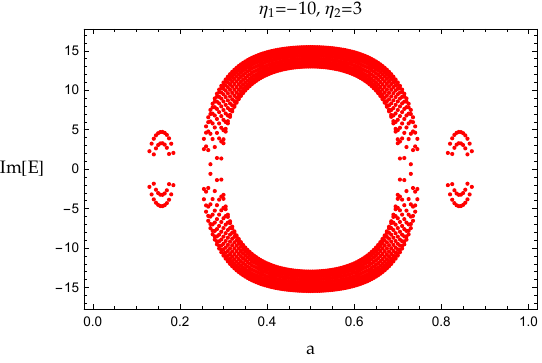}
\caption{The real and imaginary part of energy spectrum for a periodic system at the limit of  $M \rightarrow \infty$, energy spectra are determined by Eq.(\ref{spect2x}) where $\beta \in [0, \frac{2\pi}{d}]$ is quasi-momentum of electron in a periodic system.}
\label{fig5}
\end{figure*}

In order to further investigate the modification of the band structure, let's first complete the model by providing a more formal description when $\eta_1=0$ but $\eta_2 \ne 0$ (delta potentials have purely imaginary strengths). In this case, as can be seen from the characteristic determinant $D_2$ analysis, see Eq.~(\ref{d2}), the latter has no real roots for negative energy and $\mathcal{P}\mathcal{T}$
-symmetry breaks down. For the general case with $\eta_1\ne 0$, it is clear that the small values of $\eta_2$ ($\eta_2\ll \eta_1$) do not change the general shape of the band structure, but they do change the values of the energy levels. As a result, one obtains the usual band structure for the complex potential as one would
expect for a real periodic potential (\ref{spect2xa}). However, as seen from  Fig.~\ref{FIG2b} ($\eta_2=3$) and Fig.~\ref{FIG2a}  ($\eta_2=5$),    some states are already missing with increasing strength of the $\eta_2$. This applies mainly to a degenerate edge states that have separated from the upper band and move to lower band through the forbidden gap, accurately tracking the bulk bands trajectories (see panel $\eta_2=0$ in Fig.~\ref{FIG2c}). To get more insight into the mathematical structures of the eigenvalues, we focus, for clarity and simplicity, on the case of model where $a = b$ (see, for example Refs.~\cite{jons,ahmed}). For this particular case, the dispersion relation (\ref{spect2x}) can be written in the form 
\begin{equation}
\cos^2(\beta a)=\left(\cos(ka)+\frac{\eta_1}{2k}\sin(ka)\right)^2+\frac{\eta^2_2}{4k^2}\sin^2(ka) .\label{spect2xxx}
\end{equation}
First, as it is clear from Eq. (\ref{spect2xxx}), $\beta a =\pi/2$ solution is not anymore available. Second, back to the case $a\ne b$, we can state that the disappearance of some states from the band structure and the further modification of the latter with increasing strength $\eta_2$ are associated mainly with the mentioned fact of the exclusion of some antiperiodic and even non-antiperiodic solutions from the dispersion Eq. (\ref{spect2x}) (see, for example, Refs.\cite{jons,ahm}).
 
The physical meaning of the latter is completely transparent if we recall that up to a critical value $\eta_{2cr}$ of the gain and loss parameter, $\eta_2$, eigenvalues are real (see Fig. \ref{FIG2c}). At the critical or exceptional value $\eta_{2cr}$, the eigenvalues and the eigenfunctions merge. Beyond the
exceptional point, that is for $\eta_2 \geq \eta_{2cr}$, the system enters the region where some eigenvalues become complex, and complex conjugate pairs (see Fig.~\ref{FIG2a}   and Fig.~\ref{FIG2b}, right panels). The latter means that the number of real eigen-energy has decreased. The theoretical description presented above is fully confirmed by numerical simulations as illustrated in Fig.~\ref{FIG2a}   and Fig.~\ref{FIG2b}. Indeed, as can be seen from the comparison of the imaginary parts of the numerical spectra of the self-energy (red lines) in Fig.~\ref{FIG2a}   and Fig.~\ref{FIG2b}, with the increasing of $\eta_2$ the number of complex levels increases. As a result, more states begin to move out from the given region, and further modification of band structures occurs,  see e.g. Fig.~\ref{FIG2a}   and Fig.~\ref{FIG2b}.

The situation becomes even more interesting in the case of negative values of $\eta_1$, since for some values of $\eta_{2cr}$ (see below) the topological states inherent in the case $\eta_2=0$ will be pushed out. The latter means that $\eta_2$ can break the underlying symmetry of the system, including chiral or time-reversal symmetry. To study in more detail the modification of the energy spectrum in Fig.~\ref{tau1negativo2}, it is shown the band energy spectra of the finite chain for negative values of $\eta_1=-10$ for different $\eta_2$. Recall that in the case of $\eta_1 <0$ in transcendental Eqs. (\ref{C12}) and (\ref{spect2x}) $k=\sqrt {-E}$ (we are in the range $E<0$) must be replaced by $i\kappa$: as a consequence, trigonometric functions become hyperbolic. In Fig.~\ref{FIG3c}  ($\eta_2=0$) one see the existence of flat energy states in the middle of the band gap only when  $a \ge 0.5$. This asymmetry reflects the fact that the number of scatterers in the system is even (see, e.g. Refs. \cite{benj,resh,24}) and states are localized strongly over   one side of the system. The effective mass of these states is very large due the central position and, generally, they are insensitive to local perturbations, or in other words, any adiabatic deformation that respects certain symmetries of the system will not affect the existence of the symmetry-protected edge states (see, for example Refs. \cite{benj,resh,24,report}). They are classified by topological invariants that remain unchanged under continuous deformations.

It turns out, however, that the situation is somewhat more complicated in $\mathcal{P}\mathcal{T}$-symmetric  systems, since including  non-Hermitian parameter $\eta_2$ will eventually be crucial for getting insight into the physics of topological states (see, for example Ref. \cite{esaki} and references therein). The authors of \cite{esaki} concluded that zero energy modes are unstable and corresponding energy eigenvalues are not real. This conclusion is also confirmed by our analytical and numerical calculations for a one-dimensional $\mathcal{P}\mathcal{T}$-symmetric superlattice built with quantum $\delta$-wells (see below, Eq. (\ref{pot2a}) and Fig. \ref{fig4}). However, for some special cases, such as the non-Hermitian generalized off-diagonal Aubry–Andre model \cite{report}, topological edge states with real spectrum have been discovered,  provided $\eta_2$ is less than $\eta_{2cr}$. Note, that these degenerate edge states are robust against small non-Hermitian perturbations. 

Returning to the case of $\eta_2 \ne 0$, we note that a numerical analysis of 
expression (\ref{C12}) shows that as $\eta_2$ increases, the disappearance of some states from the band structure occurs even at small $\eta_2\ll \eta_{cr}$ (not shown in Fig.~\ref{tau1negativo2}).  As $\eta_2$ increases further, disappearance of more and more levels take place from the band structure (see red lines in Fig.~\ref{tau1negativo2} (a) and (b)). These lines represent the imaginary eigenvalues and are due to the fact that the spectrum is no longer real, unlike the $\eta_2=0$ case. Comparing the two reals spectrum (b) and (c) of Fig.~\ref{tau1negativo2} we note that  the real part of the spectrum is almost left unchanged except the lowest edge state energy that is pushed out from the given region (red line in Fig.~\ref{tau1negativo2} (b)). However, the band structure is drastically different for $\eta_2=3$ because of more states begin to move out, and further modification of the band structure take place. The red lines are very sensitive to the parameter $\eta_2$. We plot the real and imaginary part of the lowest edge state energy at $a = 0.8$ as a function of the $\eta_2$ in Fig. \ref{fig4}. As can be seen, actually, the complex conjugate eigenvalues appear whenever $\eta_2$ is different from zero (see right panel in Fig. \ref{fig4}).

To determine $\eta_{2cr}$ for the case of $a=b$ we solve Eqs. (\ref{C12}) and (\ref{spect2x}) simultaneously. After some algebra, we get
\begin{equation}
\eta^2_{2cr}\le\frac{{2\kappa^2}}{\sinh^2(\kappa a)}\left[ 1+\cos \left (\frac{2\pi n }{2M+1}\right) \right] ,  \ \  n=1, ..., 2M, \label{cr}
\end{equation}
with the limit 
$a^2\eta^2_{2cr}\le {2} \left [ 1+\cos \left ( \frac{2\pi n}{2M+1} \right ) \right  ]$ when $\kappa \rightarrow 0$.
The inequality (\ref{cr}) is in  a reasonable agreement with the criterion (\ref{ecr}) in the sense that it is sensitive to the parameters characterizing the system, such as number of scatters M, distance between two successive delta potentials in unit cell, etc (see below).  As the number of cells $M$ is increased, $\eta_{2cr}$ demonstrates fast oscillating behavior due to  $1+\cos \left ( \frac{2\pi n}{2M+1} \right ) $ function. The limiting condition of (\ref{cr}), when $M\rightarrow \infty$, reads ${a\eta_{2cr}}\le {2}$. With a further increase in  $\eta\geq \eta_{2cr}$ we observe drastically different shape of the energy spectrum compared to the case $\eta_2=0$ (see Fig.~\ref{FIG3a}   and Fig.~\ref{FIG3b}). This means that $\mathcal{P}\mathcal{T}$-symmetry breaks down and the eigenvalues of the bound states are complex numbers. 

At the limit of $M\rightarrow \infty$, the  $\mathcal{PT}$-symmetric system becomes periodic, and the energy spectrum is thus determined by Eq.(\ref{spect2x}), where $\beta \in [0, \frac{2\pi}{d}]$ denotes the quasi-momentum of a electron in a  periodic system. The topological edge states thus are eliminated in a periodic system, see e.g. Fig.~\ref{fig5}  compared with Fig.~\ref{FIG3a} for a finite system with hard wall boundary condition. A interesting observation is that the exceptional points  could depends heavily on $\beta$, especially for   negative energy bands, see e.g. Fig.~\ref{fig5}.

\section{Summary and outlook}
The main result of this work is that we arrive at a closed-form expression for the energy spectrum of $\mathcal{P}\mathcal{T}$-symmetric superlattice systems with two complex $\delta$-potentials in the elementary cell. It holds for both real and complex potentials, extending several well-known results in the literature and help to get even more insight into the mathematical structures of the eigenvalues.  We analyzed in detail a diatomic crystal model, varying either the scatterer distances or the potential heights. It is shown, that at a certain critical value of the imaginary part of the complex amplitude the topological states may move out. This may happened at the $\mathcal{PT}$-symmetry breaking (exceptional) points.

\acknowledgments
P.G. acknowledges support from the College of Arts and
Sciences and Faculty Research Initiative Program, Dakota
State University, Madison, SD.
 V.~G., A.~P.-G. and E.~J. would like to thank UPCT for partial
financial support through “Maria Zambrano ayudas para la recualificaci\'on del sistema universitario espa\~{n}ol 2021–2023” financed by Spanish Ministry of Universities with funds “Next Generation” of EU.  This research was supported by the National Science Foundation under Grant No. NSF PHY-2418937 and in part by the National Science Foundation under Grant No. NSF PHY1748958

\appendix

\section{Broken and unbroken $\mathcal{P}\mathcal{T}$-symmetric phases}

It is known that $\eta_2$ represents non-Hermitian degree and describes the gain and loss rate that determines the existence of a real or complex energy spectrum in $\mathcal{P}\mathcal{T}$-symmetric systems. One can show that the criterion
for the eigenvalues of $\mathcal{P}\mathcal{T}$-symmetric $S$-matrices, to be unimodular is $r_L-r_R=\pm {2i}t$  (see, e.g., Refs \cite{xy,stonea,ahmed,FS}). Depending on the physical context of the problem, a number of authors have estimated the critical value of $\eta_{2cr}$ either based on the above criterion (see, e.g., \cite{stonea}) or by diagonalizing the Hamiltonian and obtaining the energy spectrum numerically (see, e.g., \cite{japan}), or by relating the problem to an solvable effective two-state Hamiltonian problem, that holds only in the thermodynamic limit (see, e.g., \cite{hay,japan1}).  At this limit, near the critical point, all microscopic details of the system will eventually become irrelevant, contrary to what would be expected from the finite system. Regarding the latter, some analytical/numerical results for the critical frequency at which the symmetry breaking phase occurs in a complex index dielectric slab or in an arbitrary one dimensional finite periodic scatterer or predicting the $\mathcal{P}\mathcal{T}$ -symmetry breaking point for edge states in finite lattices are published in \cite{stonea,FS,hay}. Based on a tight-biding model and assuming $\eta_1=0$, it was shown that the transition from weak  non-Hermiticity to the regime of strong non-Hermiticity is controlled by the ratio $\frac{\eta_2}{2t_h}$ ($t_h$ is the amplitude of the jump between
two sites and a gain/loss pair controlled by the $\mathcal{P}\mathcal{T}$ $\eta_2$ parameter) (see, e.g., Refs. \cite{japan2,xx,song}). The behavior of the transmission probability is strongly non-Hermitian in the regime of weak non-Hermiticity with divergent peaks when $\frac{\eta_2}{2t_h}<1$ and is almost Hermitian in the regime of strong non-Hermiticity, $\frac{\eta_2}{2t_h}>1$. For a simplified continuum model consisting of two complex delta potentials, it was found that divergent peaks of the transmission and reflection coefficients appear in the range $\frac{\eta_2}{2k}<1$ (see, e.g., Refs. \cite{japan2,6,peng1,peng2}). 

In what follows, we will rewrite the above criterion $r_L-r_R=\pm {2i}t$ for broken-symmetry phase/$\mathcal{P}\mathcal{T}$-symmetric phase in a slightly different, but completely equivalent form. This consists of mapping the mentioned criterion onto a one-dimensional partial differential equation, leaving only one parameter $t$. We will see that the result obtained with the new criterion is consistent with continuum results obtained elsewhere using single-particle scattering from an absorbing delta-potential (see, e.g., Refs. \cite{muga,japan,FS,japan1}).

To this end, we focus on the case $Z_1=Z^*_N=\eta_1+i\eta_2$ and note that using the expressions (\ref{rL}) and (\ref{rR}) for $r_L$ and $r_R$ the criterion $r_L-r_R=\pm {2i}t$ can be rewritten in equivalent form
\begin{equation}
\frac{\partial}{\partial Z_N}\frac{1}{t}\bigg|_{cr}-\frac{\partial}{\partial Z_1}\frac{1}{t}\bigg|_{cr}=\pm \frac{1}{k} 
. \label{r_L2b}
\end{equation}
Using the chain rule and rewriting the above equation in terms of partial derivatives with respect to $\eta_1$ and $\eta_2$ we arrived at:
\begin{equation}
i\frac{\partial}{\partial \eta_2}\frac{1}{t}\bigg|_{cr}=\pm\frac{1}{k} .
\end{equation}
Integrating both sides of the equation and assuming that an arbitrary function representing the constant of integration with respect to $\eta_2$ is not zero, say  $\frac{1}{t_0}$, the solution can be presented the form
\begin{equation}
\frac{1}{t}\bigg|_{cr}=\mp \frac{i\eta_{2cr}}{k}+\frac{1}{t_0}.\label{ecr}
\end{equation}
The above criterion (\ref{ecr}) for $1/t_{cr}$ depends solely by $\eta_{2cr}/k$ (first term) and  on the form of the 1D shape of the potential (second term).  The particular solution, when $\frac{1}{t_0}$ is zero, nicely reproduces the transmission amplitude as function of the strength of the imaginary part of complex $\delta$ function at an exceptional point, that is around $\eta_2 \approx k$. This also allows us to understand, at least at a qualitative level, why diverging peaks of transmission and reflection coefficients appear in the range $\frac{\eta_2}{2k}<1$ (see, e.g., Refs. \cite{japan2,peng1,peng2}). It is easy to get convinced that if the constant $\frac{1}{t_0}$ of integration over $\eta_2$ is not equal to zero, then the explicit value of $\eta_{cr}$ is generally very complex and depends on the shape of the potential.

\bibliography{ALL-REF.bib}

\begin{thebibliography}{48}%
\makeatletter
\providecommand \@ifxundefined [1]{%
 \@ifx{#1\undefined}
}%
\providecommand \@ifnum [1]{%
 \ifnum #1\expandafter \@firstoftwo
 \else \expandafter \@secondoftwo
 \fi
}%
\providecommand \@ifx [1]{%
 \ifx #1\expandafter \@firstoftwo
 \else \expandafter \@secondoftwo
 \fi
}%
\providecommand \natexlab [1]{#1}%
\providecommand \enquote  [1]{``#1''}%
\providecommand \bibnamefont  [1]{#1}%
\providecommand \bibfnamefont [1]{#1}%
\providecommand \citenamefont [1]{#1}%
\providecommand \href@noop [0]{\@secondoftwo}%
\providecommand \href [0]{\begingroup \@sanitize@url \@href}%
\providecommand \@href[1]{\@@startlink{#1}\@@href}%
\providecommand \@@href[1]{\endgroup#1\@@endlink}%
\providecommand \@sanitize@url [0]{\catcode `\\12\catcode `\$12\catcode
  `\&12\catcode `\#12\catcode `\^12\catcode `\_12\catcode `\%12\relax}%
\providecommand \@@startlink[1]{}%
\providecommand \@@endlink[0]{}%
\providecommand \url  [0]{\begingroup\@sanitize@url \@url }%
\providecommand \@url [1]{\endgroup\@href {#1}{\urlprefix }}%
\providecommand \urlprefix  [0]{URL }%
\providecommand \Eprint [0]{\href }%
\providecommand \doibase [0]{http://dx.doi.org/}%
\providecommand \selectlanguage [0]{\@gobble}%
\providecommand \bibinfo  [0]{\@secondoftwo}%
\providecommand \bibfield  [0]{\@secondoftwo}%
\providecommand \translation [1]{[#1]}%
\providecommand \BibitemOpen [0]{}%
\providecommand \bibitemStop [0]{}%
\providecommand \bibitemNoStop [0]{.\EOS\space}%
\providecommand \EOS [0]{\spacefactor3000\relax}%
\providecommand \BibitemShut  [1]{\csname bibitem#1\endcsname}%
\let\auto@bib@innerbib\@empty
\bibitem [{\citenamefont {Bender}\ and\ \citenamefont {Boettcher}(1998)}]{1a}%
  \BibitemOpen
  \bibfield  {author} {\bibinfo {author} {\bibfnamefont {Carl~M.}\ \bibnamefont
  {Bender}}\ and\ \bibinfo {author} {\bibfnamefont {Stefan}\ \bibnamefont
  {Boettcher}},\ }\bibfield  {title} {\enquote {\bibinfo {title} {Real spectra
  in non-hermitian hamiltonians having $\textsc{P}\textsc{T}$ symmetry},}\
  }\href {\doibase 10.1103/PhysRevLett.80.5243} {\bibfield  {journal} {\bibinfo
   {journal} {Phys. Rev. Lett.}\ }\textbf {\bibinfo {volume} {80}},\ \bibinfo
  {pages} {5243--5246} (\bibinfo {year} {1998})}\BibitemShut {NoStop}%
\bibitem [{\citenamefont {Jones}(1999)}]{jons}%
  \BibitemOpen
  \bibfield  {author} {\bibinfo {author} {\bibfnamefont {H.F}\ \bibnamefont
  {Jones}},\ }\bibfield  {title} {\enquote {\bibinfo {title} {The energy
  spectrum of complex periodic potentials of the kronig–penney type},}\
  }\href {\doibase https://doi.org/10.1016/S0375-9601(99)00672-6} {\bibfield
  {journal} {\bibinfo  {journal} {Physics Letters A}\ }\textbf {\bibinfo
  {volume} {262}},\ \bibinfo {pages} {242--244} (\bibinfo {year}
  {1999})}\BibitemShut {NoStop}%
\bibitem [{\citenamefont {Crassee}\ \emph {et~al.}(2011)\citenamefont
  {Crassee}, \citenamefont {Levallois}, \citenamefont {Walter}, \citenamefont
  {Ostler}, \citenamefont {Bostwick}, \citenamefont {Rotenberg}, \citenamefont
  {Seyller}, \citenamefont {van~der Marel},\ and\ \citenamefont
  {Kuzmenko}}]{2a}%
  \BibitemOpen
  \bibfield  {author} {\bibinfo {author} {\bibfnamefont {Iris}\ \bibnamefont
  {Crassee}}, \bibinfo {author} {\bibfnamefont {Julien}\ \bibnamefont
  {Levallois}}, \bibinfo {author} {\bibfnamefont {Andrew~L.}\ \bibnamefont
  {Walter}}, \bibinfo {author} {\bibfnamefont {Markus}\ \bibnamefont {Ostler}},
  \bibinfo {author} {\bibfnamefont {Aaron}\ \bibnamefont {Bostwick}}, \bibinfo
  {author} {\bibfnamefont {Eli}\ \bibnamefont {Rotenberg}}, \bibinfo {author}
  {\bibfnamefont {Thomas}\ \bibnamefont {Seyller}}, \bibinfo {author}
  {\bibfnamefont {Dirk}\ \bibnamefont {van~der Marel}}, \ and\ \bibinfo
  {author} {\bibfnamefont {Alexey~B.}\ \bibnamefont {Kuzmenko}},\ }\bibfield
  {title} {\enquote {\bibinfo {title} {Giant faraday rotation in single- and
  multilayer graphene},}\ }\href {\doibase 10.1038/nphys1816} {\bibfield
  {journal} {\bibinfo  {journal} {Nature Physics}\ }\textbf {\bibinfo {volume}
  {7}},\ \bibinfo {pages} {48--51} (\bibinfo {year} {2011})}\BibitemShut
  {NoStop}%
\bibitem [{\citenamefont {Guo}\ \emph {et~al.}(2009)\citenamefont {Guo},
  \citenamefont {Salamo}, \citenamefont {Duchesne}, \citenamefont {Morandotti},
  \citenamefont {Volatier-Ravat}, \citenamefont {Aimez}, \citenamefont
  {Siviloglou},\ and\ \citenamefont {Christodoulides}}]{3}%
  \BibitemOpen
  \bibfield  {author} {\bibinfo {author} {\bibfnamefont {A.}~\bibnamefont
  {Guo}}, \bibinfo {author} {\bibfnamefont {G.~J.}\ \bibnamefont {Salamo}},
  \bibinfo {author} {\bibfnamefont {D.}~\bibnamefont {Duchesne}}, \bibinfo
  {author} {\bibfnamefont {R.}~\bibnamefont {Morandotti}}, \bibinfo {author}
  {\bibfnamefont {M.}~\bibnamefont {Volatier-Ravat}}, \bibinfo {author}
  {\bibfnamefont {V.}~\bibnamefont {Aimez}}, \bibinfo {author} {\bibfnamefont
  {G.~A.}\ \bibnamefont {Siviloglou}}, \ and\ \bibinfo {author} {\bibfnamefont
  {D.~N.}\ \bibnamefont {Christodoulides}},\ }\bibfield  {title} {\enquote
  {\bibinfo {title} {Observation of $\mathcal{PT}$-symmetry breaking in complex
  optical potentials},}\ }\href {\doibase 10.1103/PhysRevLett.103.093902}
  {\bibfield  {journal} {\bibinfo  {journal} {Phys. Rev. Lett.}\ }\textbf
  {\bibinfo {volume} {103}},\ \bibinfo {pages} {093902} (\bibinfo {year}
  {2009})}\BibitemShut {NoStop}%
\bibitem [{\citenamefont {Zyablovsky}\ \emph {et~al.}(2014)\citenamefont
  {Zyablovsky}, \citenamefont {Vinogradov}, \citenamefont {Pukhov},
  \citenamefont {Dorofeenko},\ and\ \citenamefont {Lisyansky}}]{4}%
  \BibitemOpen
  \bibfield  {author} {\bibinfo {author} {\bibfnamefont {A~A}\ \bibnamefont
  {Zyablovsky}}, \bibinfo {author} {\bibfnamefont {A~P}\ \bibnamefont
  {Vinogradov}}, \bibinfo {author} {\bibfnamefont {A~A}\ \bibnamefont
  {Pukhov}}, \bibinfo {author} {\bibfnamefont {A~V}\ \bibnamefont
  {Dorofeenko}}, \ and\ \bibinfo {author} {\bibfnamefont {A~A}\ \bibnamefont
  {Lisyansky}},\ }\bibfield  {title} {\enquote {\bibinfo {title} {Pt-symmetry
  in optics},}\ }\href {\doibase 10.3367/UFNe.0184.201411b.1177} {\bibfield
  {journal} {\bibinfo  {journal} {Physics-Uspekhi}\ }\textbf {\bibinfo {volume}
  {57}},\ \bibinfo {pages} {1063} (\bibinfo {year} {2014})}\BibitemShut
  {NoStop}%
\bibitem [{\citenamefont {Makris}\ \emph {et~al.}(2008)\citenamefont {Makris},
  \citenamefont {El-Ganainy}, \citenamefont {Christodoulides},\ and\
  \citenamefont {Musslimani}}]{5}%
  \BibitemOpen
  \bibfield  {author} {\bibinfo {author} {\bibfnamefont {K.~G.}\ \bibnamefont
  {Makris}}, \bibinfo {author} {\bibfnamefont {R.}~\bibnamefont {El-Ganainy}},
  \bibinfo {author} {\bibfnamefont {D.~N.}\ \bibnamefont {Christodoulides}}, \
  and\ \bibinfo {author} {\bibfnamefont {Z.~H.}\ \bibnamefont {Musslimani}},\
  }\bibfield  {title} {\enquote {\bibinfo {title} {Beam dynamics in
  $\mathcal{P}\mathcal{T}$ symmetric optical lattices},}\ }\href {\doibase
  10.1103/PhysRevLett.100.103904} {\bibfield  {journal} {\bibinfo  {journal}
  {Phys. Rev. Lett.}\ }\textbf {\bibinfo {volume} {100}},\ \bibinfo {pages}
  {103904} (\bibinfo {year} {2008})}\BibitemShut {NoStop}%
\bibitem [{\citenamefont {Dani}\ \emph {et~al.}(2011)\citenamefont {Dani},
  \citenamefont {Wang}, \citenamefont {Bossmann}, \citenamefont {Wysin},\ and\
  \citenamefont {Chikan}}]{1}%
  \BibitemOpen
  \bibfield  {author} {\bibinfo {author} {\bibfnamefont {Raj~Kumar}\
  \bibnamefont {Dani}}, \bibinfo {author} {\bibfnamefont {Hongwang}\
  \bibnamefont {Wang}}, \bibinfo {author} {\bibfnamefont {Stefan~H.}\
  \bibnamefont {Bossmann}}, \bibinfo {author} {\bibfnamefont {Gary}\
  \bibnamefont {Wysin}}, \ and\ \bibinfo {author} {\bibfnamefont {Viktor}\
  \bibnamefont {Chikan}},\ }\bibfield  {title} {\enquote {\bibinfo {title}
  {Faraday rotation enhancement of gold coated fe2o3 nanoparticles: Comparison
  of experiment and theory},}\ }\href {\doibase 10.1063/1.3665138} {\bibfield
  {journal} {\bibinfo  {journal} {The Journal of Chemical Physics}\ }\textbf
  {\bibinfo {volume} {135}},\ \bibinfo {pages} {224502} (\bibinfo {year}
  {2011})},\ \Eprint
  {http://arxiv.org/abs/https://pubs.aip.org/aip/jcp/article-pdf/doi/10.1063/1.3665138/15445749/224502\_1\_online.pdf}
  {https://pubs.aip.org/aip/jcp/article-pdf/doi/10.1063/1.3665138/15445749/224502\_1\_online.pdf}
  \BibitemShut {NoStop}%
\bibitem [{\citenamefont {El-Ganainy}\ \emph {et~al.}(2018)\citenamefont
  {El-Ganainy}, \citenamefont {Makris}, \citenamefont {Khajavikhan},
  \citenamefont {Musslimani}, \citenamefont {Rotter},\ and\ \citenamefont
  {Christodoulides}}]{2}%
  \BibitemOpen
  \bibfield  {author} {\bibinfo {author} {\bibfnamefont {Ramy}\ \bibnamefont
  {El-Ganainy}}, \bibinfo {author} {\bibfnamefont {Konstantinos~G.}\
  \bibnamefont {Makris}}, \bibinfo {author} {\bibfnamefont {Mercedeh}\
  \bibnamefont {Khajavikhan}}, \bibinfo {author} {\bibfnamefont {Ziad~H.}\
  \bibnamefont {Musslimani}}, \bibinfo {author} {\bibfnamefont {Stefan}\
  \bibnamefont {Rotter}}, \ and\ \bibinfo {author} {\bibfnamefont
  {Demetrios~N.}\ \bibnamefont {Christodoulides}},\ }\bibfield  {title}
  {\enquote {\bibinfo {title} {Non-hermitian physics and pt symmetry},}\ }\href
  {\doibase 10.1038/nphys4323} {\bibfield  {journal} {\bibinfo  {journal}
  {Nature Physics}\ }\textbf {\bibinfo {volume} {14}},\ \bibinfo {pages}
  {11--19} (\bibinfo {year} {2018})}\BibitemShut {NoStop}%
\bibitem [{\citenamefont {Burke}\ \emph
  {et~al.}(2020{\natexlab{a}})\citenamefont {Burke}, \citenamefont {Wiersig},\
  and\ \citenamefont {Haque}}]{6}%
  \BibitemOpen
  \bibfield  {author} {\bibinfo {author} {\bibfnamefont {Phillip~C.}\
  \bibnamefont {Burke}}, \bibinfo {author} {\bibfnamefont {Jan}\ \bibnamefont
  {Wiersig}}, \ and\ \bibinfo {author} {\bibfnamefont {Masudul}\ \bibnamefont
  {Haque}},\ }\bibfield  {title} {\enquote {\bibinfo {title} {Non-hermitian
  scattering on a tight-binding lattice},}\ }\href {\doibase
  10.1103/PhysRevA.102.012212} {\bibfield  {journal} {\bibinfo  {journal}
  {Phys. Rev. A}\ }\textbf {\bibinfo {volume} {102}},\ \bibinfo {pages}
  {012212} (\bibinfo {year} {2020}{\natexlab{a}})}\BibitemShut {NoStop}%
\bibitem [{\citenamefont {Huang}\ \emph {et~al.}(2021)\citenamefont {Huang},
  \citenamefont {He}, \citenamefont {Li}, \citenamefont {Wu}, \citenamefont
  {Zhang}, \citenamefont {Jin},\ and\ \citenamefont {Gong}}]{7}%
  \BibitemOpen
  \bibfield  {author} {\bibinfo {author} {\bibfnamefont {Piao-Piao}\
  \bibnamefont {Huang}}, \bibinfo {author} {\bibfnamefont {Jing}\ \bibnamefont
  {He}}, \bibinfo {author} {\bibfnamefont {Jia-Rui}\ \bibnamefont {Li}},
  \bibinfo {author} {\bibfnamefont {Hai-Na}\ \bibnamefont {Wu}}, \bibinfo
  {author} {\bibfnamefont {Lian-Lian}\ \bibnamefont {Zhang}}, \bibinfo {author}
  {\bibfnamefont {Zhao}\ \bibnamefont {Jin}}, \ and\ \bibinfo {author}
  {\bibfnamefont {Wei-Jiang}\ \bibnamefont {Gong}},\ }\bibfield  {title}
  {\enquote {\bibinfo {title} {Transmission through a one-dimensional photonic
  lattice modulated by the side-coupled pt-symmetric non-hermitian
  su--schrieffer--heeger chain},}\ }\href {\doibase 10.1364/JOSAB.411989}
  {\bibfield  {journal} {\bibinfo  {journal} {J. Opt. Soc. Am. B}\ }\textbf
  {\bibinfo {volume} {38}},\ \bibinfo {pages} {1331--1340} (\bibinfo {year}
  {2021})}\BibitemShut {NoStop}%
\bibitem [{\citenamefont {Zheng}\ \emph {et~al.}(2019)\citenamefont {Zheng},
  \citenamefont {Yang}, \citenamefont {Deng},\ and\ \citenamefont {Liu}}]{xx}%
  \BibitemOpen
  \bibfield  {author} {\bibinfo {author} {\bibfnamefont {Jian}\ \bibnamefont
  {Zheng}}, \bibinfo {author} {\bibfnamefont {Xiangbo}\ \bibnamefont {Yang}},
  \bibinfo {author} {\bibfnamefont {Dongmei}\ \bibnamefont {Deng}}, \ and\
  \bibinfo {author} {\bibfnamefont {Hongzhan}\ \bibnamefont {Liu}},\ }\bibfield
   {title} {\enquote {\bibinfo {title} {Singular properties generated by finite
  periodic pt-symmetric optical waveguide network},}\ }\href {\doibase
  10.1364/OE.27.001538} {\bibfield  {journal} {\bibinfo  {journal} {Opt.
  Express}\ }\textbf {\bibinfo {volume} {27}},\ \bibinfo {pages} {1538--1552}
  (\bibinfo {year} {2019})}\BibitemShut {NoStop}%
\bibitem [{\citenamefont {Ge}\ \emph {et~al.}(2012)\citenamefont {Ge},
  \citenamefont {Chong},\ and\ \citenamefont {Stone}}]{xy}%
  \BibitemOpen
  \bibfield  {author} {\bibinfo {author} {\bibfnamefont {Li}~\bibnamefont
  {Ge}}, \bibinfo {author} {\bibfnamefont {Y.~D.}\ \bibnamefont {Chong}}, \
  and\ \bibinfo {author} {\bibfnamefont {A.~D.}\ \bibnamefont {Stone}},\
  }\bibfield  {title} {\enquote {\bibinfo {title} {Conservation relations and
  anisotropic transmission resonances in one-dimensional
  $\mathcal{PT}$-symmetric photonic heterostructures},}\ }\href {\doibase
  10.1103/PhysRevA.85.023802} {\bibfield  {journal} {\bibinfo  {journal} {Phys.
  Rev. A}\ }\textbf {\bibinfo {volume} {85}},\ \bibinfo {pages} {023802}
  (\bibinfo {year} {2012})}\BibitemShut {NoStop}%
\bibitem [{\citenamefont {Basiri}\ \emph {et~al.}(2015)\citenamefont {Basiri},
  \citenamefont {Vitebskiy},\ and\ \citenamefont {Kottos}}]{zz1}%
  \BibitemOpen
  \bibfield  {author} {\bibinfo {author} {\bibfnamefont {A.}~\bibnamefont
  {Basiri}}, \bibinfo {author} {\bibfnamefont {I.}~\bibnamefont {Vitebskiy}}, \
  and\ \bibinfo {author} {\bibfnamefont {T.}~\bibnamefont {Kottos}},\
  }\bibfield  {title} {\enquote {\bibinfo {title} {Light scattering in
  pseudopassive media with uniformly balanced gain and loss},}\ }\href
  {\doibase 10.1103/PhysRevA.91.063843} {\bibfield  {journal} {\bibinfo
  {journal} {Phys. Rev. A}\ }\textbf {\bibinfo {volume} {91}},\ \bibinfo
  {pages} {063843} (\bibinfo {year} {2015})}\BibitemShut {NoStop}%
\bibitem [{\citenamefont {Guo}\ and\ \citenamefont
  {Gasparian}(2022)}]{Guo:2022row}%
  \BibitemOpen
  \bibfield  {author} {\bibinfo {author} {\bibfnamefont {Peng}\ \bibnamefont
  {Guo}}\ and\ \bibinfo {author} {\bibfnamefont {Vladimir}\ \bibnamefont
  {Gasparian}},\ }\bibfield  {title} {\enquote {\bibinfo {title} {{Friedel
  formula and Krein's theorem in complex potential scattering theory}},}\
  }\href {\doibase 10.1103/PhysRevResearch.4.023083} {\bibfield  {journal}
  {\bibinfo  {journal} {Phys. Rev. Res.}\ }\textbf {\bibinfo {volume} {4}},\
  \bibinfo {pages} {023083} (\bibinfo {year} {2022})},\ \Eprint
  {http://arxiv.org/abs/2202.12465} {arXiv:2202.12465 [cond-mat.other]}
  \BibitemShut {NoStop}%
\bibitem [{\citenamefont {Guo}\ \emph {et~al.}(2023)\citenamefont {Guo},
  \citenamefont {Gasparian}, \citenamefont {J\'odar},\ and\ \citenamefont
  {Wisehart}}]{peng1}%
  \BibitemOpen
  \bibfield  {author} {\bibinfo {author} {\bibfnamefont {Peng}\ \bibnamefont
  {Guo}}, \bibinfo {author} {\bibfnamefont {Vladimir}\ \bibnamefont
  {Gasparian}}, \bibinfo {author} {\bibfnamefont {Esther}\ \bibnamefont
  {J\'odar}}, \ and\ \bibinfo {author} {\bibfnamefont {Christopher}\
  \bibnamefont {Wisehart}},\ }\bibfield  {title} {\enquote {\bibinfo {title}
  {Tunneling time in $\mathcal{PT}$-symmetric systems},}\ }\href {\doibase
  10.1103/PhysRevA.107.032210} {\bibfield  {journal} {\bibinfo  {journal}
  {Phys. Rev. A}\ }\textbf {\bibinfo {volume} {107}},\ \bibinfo {pages}
  {032210} (\bibinfo {year} {2023})}\BibitemShut {NoStop}%
\bibitem [{\citenamefont {Gasparian}\ \emph {et~al.}(2023)\citenamefont
  {Gasparian}, \citenamefont {Guo}, \citenamefont {Pérez-Garrido},\ and\
  \citenamefont {Jódar}}]{peng2}%
  \BibitemOpen
  \bibfield  {author} {\bibinfo {author} {\bibfnamefont {Vladimir}\
  \bibnamefont {Gasparian}}, \bibinfo {author} {\bibfnamefont {Peng}\
  \bibnamefont {Guo}}, \bibinfo {author} {\bibfnamefont {Antonio}\ \bibnamefont
  {Pérez-Garrido}}, \ and\ \bibinfo {author} {\bibfnamefont {Esther}\
  \bibnamefont {Jódar}},\ }\bibfield  {title} {\enquote {\bibinfo {title}
  {Tunneling time and faraday/kerr effects in $\mathcal{PT}$\text{-symmetric}
  systems},}\ }\href {\doibase 10.1209/0295-5075/acf59e} {\bibfield  {journal}
  {\bibinfo  {journal} {Europhysics Letters}\ }\textbf {\bibinfo {volume}
  {143}},\ \bibinfo {pages} {66001} (\bibinfo {year} {2023})}\BibitemShut
  {NoStop}%
\bibitem [{\citenamefont {Guo}\ \emph {et~al.}(2024)\citenamefont {Guo},
  \citenamefont {Gasparian}, \citenamefont {P\'erez-Garrido},\ and\
  \citenamefont {J\'odar}}]{Guo:2024bar}%
  \BibitemOpen
  \bibfield  {author} {\bibinfo {author} {\bibfnamefont {Peng}\ \bibnamefont
  {Guo}}, \bibinfo {author} {\bibfnamefont {Vladimir}\ \bibnamefont
  {Gasparian}}, \bibinfo {author} {\bibfnamefont {Antonio}\ \bibnamefont
  {P\'erez-Garrido}}, \ and\ \bibinfo {author} {\bibfnamefont {Esther}\
  \bibnamefont {J\'odar}},\ }\bibfield  {title} {\enquote {\bibinfo {title}
  {{Tunneling time in coupled-channel systems}},}\ }\href {\doibase
  10.1103/PhysRevResearch.6.043032} {\bibfield  {journal} {\bibinfo  {journal}
  {Phys. Rev. Res.}\ }\textbf {\bibinfo {volume} {6}},\ \bibinfo {pages}
  {043032} (\bibinfo {year} {2024})},\ \Eprint
  {http://arxiv.org/abs/2407.17981} {arXiv:2407.17981 [cond-mat.other]}
  \BibitemShut {NoStop}%
\bibitem [{\citenamefont {Gasparian}\ \emph {et~al.}(2022)\citenamefont
  {Gasparian}, \citenamefont {Guo},\ and\ \citenamefont
  {J\'odar}}]{Gasparian:2022acf}%
  \BibitemOpen
  \bibfield  {author} {\bibinfo {author} {\bibfnamefont {Vladimir}\
  \bibnamefont {Gasparian}}, \bibinfo {author} {\bibfnamefont {Peng}\
  \bibnamefont {Guo}}, \ and\ \bibinfo {author} {\bibfnamefont {Esther}\
  \bibnamefont {J\'odar}},\ }\bibfield  {title} {\enquote {\bibinfo {title}
  {{Anomalous Faraday effect in a PT-symmetric dielectric slab}},}\ }\href
  {\doibase 10.1016/j.physleta.2022.128473} {\bibfield  {journal} {\bibinfo
  {journal} {Phys. Lett. A}\ }\textbf {\bibinfo {volume} {453}},\ \bibinfo
  {pages} {128473} (\bibinfo {year} {2022})},\ \Eprint
  {http://arxiv.org/abs/2205.09871} {arXiv:2205.09871 [cond-mat.mes-hall]}
  \BibitemShut {NoStop}%
\bibitem [{\citenamefont {Perez-Garrido}\ \emph {et~al.}(2023)\citenamefont
  {Perez-Garrido}, \citenamefont {Guo}, \citenamefont {Gasparian},\ and\
  \citenamefont {J\'odar}}]{Guo1}%
  \BibitemOpen
  \bibfield  {author} {\bibinfo {author} {\bibfnamefont {Antonio}\ \bibnamefont
  {Perez-Garrido}}, \bibinfo {author} {\bibfnamefont {Peng}\ \bibnamefont
  {Guo}}, \bibinfo {author} {\bibfnamefont {Vladimir}\ \bibnamefont
  {Gasparian}}, \ and\ \bibinfo {author} {\bibfnamefont {Esther}\ \bibnamefont
  {J\'odar}},\ }\bibfield  {title} {\enquote {\bibinfo {title} {Polar
  magneto-optic kerr and faraday effects in finite periodic
  $\mathcal{PT}$-symmetric systems},}\ }\href {\doibase
  10.1103/PhysRevA.107.053504} {\bibfield  {journal} {\bibinfo  {journal}
  {Phys. Rev. A}\ }\textbf {\bibinfo {volume} {107}},\ \bibinfo {pages}
  {053504} (\bibinfo {year} {2023})}\BibitemShut {NoStop}%
\bibitem [{\citenamefont {Gasparian}\ \emph {et~al.}(1988)\citenamefont
  {Gasparian}, \citenamefont {Altshuler}, \citenamefont {Aronov},\ and\
  \citenamefont {Kasamanian}}]{GA88}%
  \BibitemOpen
  \bibfield  {author} {\bibinfo {author} {\bibfnamefont {V.M.}\ \bibnamefont
  {Gasparian}}, \bibinfo {author} {\bibfnamefont {B.L.}\ \bibnamefont
  {Altshuler}}, \bibinfo {author} {\bibfnamefont {A.G.}\ \bibnamefont
  {Aronov}}, \ and\ \bibinfo {author} {\bibfnamefont {Z.A.}\ \bibnamefont
  {Kasamanian}},\ }\bibfield  {title} {\enquote {\bibinfo {title} {Resistance
  of one-dimensional chains in kronig-penny-like models},}\ }\href {\doibase
  https://doi.org/10.1016/0375-9601(88)90284-8} {\bibfield  {journal} {\bibinfo
   {journal} {Physics Letters A}\ }\textbf {\bibinfo {volume} {132}},\ \bibinfo
  {pages} {201--205} (\bibinfo {year} {1988})}\BibitemShut {NoStop}%
\bibitem [{\citenamefont {Gasparian}\ and\ \citenamefont
  {Pollak}(1993)}]{Pollak}%
  \BibitemOpen
  \bibfield  {author} {\bibinfo {author} {\bibfnamefont {V.}~\bibnamefont
  {Gasparian}}\ and\ \bibinfo {author} {\bibfnamefont {M.}~\bibnamefont
  {Pollak}},\ }\bibfield  {title} {\enquote {\bibinfo {title}
  {B\"uttiker-landauer characteristic barrier-interaction times for
  one-dimensional random layered systems},}\ }\href {\doibase
  10.1103/PhysRevB.47.2038} {\bibfield  {journal} {\bibinfo  {journal} {Phys.
  Rev. B}\ }\textbf {\bibinfo {volume} {47}},\ \bibinfo {pages} {2038--2041}
  (\bibinfo {year} {1993})}\BibitemShut {NoStop}%
\bibitem [{\citenamefont {Gasparian}\ \emph
  {et~al.}(1997{\natexlab{a}})\citenamefont {Gasparian}, \citenamefont
  {Gummich}, \citenamefont {Jódar}, \citenamefont {Ruiz},\ and\ \citenamefont
  {Ortuño}}]{gas1}%
  \BibitemOpen
  \bibfield  {author} {\bibinfo {author} {\bibfnamefont {V.}~\bibnamefont
  {Gasparian}}, \bibinfo {author} {\bibfnamefont {Ute}\ \bibnamefont
  {Gummich}}, \bibinfo {author} {\bibfnamefont {E.}~\bibnamefont {Jódar}},
  \bibinfo {author} {\bibfnamefont {J.}~\bibnamefont {Ruiz}}, \ and\ \bibinfo
  {author} {\bibfnamefont {M.}~\bibnamefont {Ortuño}},\ }\bibfield  {title}
  {\enquote {\bibinfo {title} {Tunneling and dwell time for one-dimensional
  generalized kronig-penney model},}\ }\href {\doibase
  https://doi.org/10.1016/S0921-4526(96)01036-8} {\bibfield  {journal}
  {\bibinfo  {journal} {Physica B: Condensed Matter}\ }\textbf {\bibinfo
  {volume} {233}},\ \bibinfo {pages} {72--77} (\bibinfo {year}
  {1997}{\natexlab{a}})}\BibitemShut {NoStop}%
\bibitem [{\citenamefont {Tzortzakakis}\ \emph {et~al.}(2022)\citenamefont
  {Tzortzakakis}, \citenamefont {Katsaris}, \citenamefont {Palaiodimopoulos},
  \citenamefont {Kalozoumis}, \citenamefont {Theocharis}, \citenamefont
  {Diakonos},\ and\ \citenamefont {Petrosyan}}]{hay}%
  \BibitemOpen
  \bibfield  {author} {\bibinfo {author} {\bibfnamefont {A.~F.}\ \bibnamefont
  {Tzortzakakis}}, \bibinfo {author} {\bibfnamefont {A.}~\bibnamefont
  {Katsaris}}, \bibinfo {author} {\bibfnamefont {N.~E.}\ \bibnamefont
  {Palaiodimopoulos}}, \bibinfo {author} {\bibfnamefont {P.~A.}\ \bibnamefont
  {Kalozoumis}}, \bibinfo {author} {\bibfnamefont {G.}~\bibnamefont
  {Theocharis}}, \bibinfo {author} {\bibfnamefont {F.~K.}\ \bibnamefont
  {Diakonos}}, \ and\ \bibinfo {author} {\bibfnamefont {D.}~\bibnamefont
  {Petrosyan}},\ }\bibfield  {title} {\enquote {\bibinfo {title} {Topological
  edge states of the $\mathcal{PT}$-symmetric su-schrieffer-heeger model: An
  effective two-state description},}\ }\href {\doibase
  10.1103/PhysRevA.106.023513} {\bibfield  {journal} {\bibinfo  {journal}
  {Phys. Rev. A}\ }\textbf {\bibinfo {volume} {106}},\ \bibinfo {pages}
  {023513} (\bibinfo {year} {2022})}\BibitemShut {NoStop}%
\bibitem [{\citenamefont {{Jin}}(2017)}]{chin}%
  \BibitemOpen
  \bibfield  {author} {\bibinfo {author} {\bibfnamefont {L.}~\bibnamefont
  {{Jin}}},\ }\bibfield  {title} {\enquote {\bibinfo {title} {{Topological
  phases and edge states in a non-Hermitian trimerized optical lattice}},}\
  }\href {\doibase 10.1103/PhysRevA.96.032103} {\bibfield  {journal} {\bibinfo
  {journal} {\pra}\ }\textbf {\bibinfo {volume} {96}},\ \bibinfo {eid} {032103}
  (\bibinfo {year} {2017})},\ \Eprint {http://arxiv.org/abs/1803.06672}
  {arXiv:1803.06672 [cond-mat.mes-hall]} \BibitemShut {NoStop}%
\bibitem [{\citenamefont {Kasamanian}(1972)}]{zatik}%
  \BibitemOpen
  \bibfield  {author} {\bibinfo {author} {\bibfnamefont {Z.A.}\ \bibnamefont
  {Kasamanian}},\ }\bibfield  {title} {\enquote {\bibinfo {title} {{On the
  theory of impurity levels}},}\ }\href@noop {} {\bibfield  {journal} {\bibinfo
   {journal} {Sov. Phys. JETP}\ }\textbf {\bibinfo {volume} {34}},\ \bibinfo
  {pages} {648} (\bibinfo {year} {1972})},\ \bibinfo {note} {[Zh. Eksp. Teor.
  Fiz. 61, 1215-1220(1971)]}\BibitemShut {NoStop}%
\bibitem [{\citenamefont {Economou}(2006)}]{eco}%
  \BibitemOpen
  \bibfield  {author} {\bibinfo {author} {\bibfnamefont {E.N.}\ \bibnamefont
  {Economou}},\ }\href {https://books.google.com/books?id=HdJDAAAAQBAJ} {\emph
  {\bibinfo {title} {Green's Functions in Quantum Physics}}},\ Springer Series
  in Solid-State Sciences\ (\bibinfo  {publisher} {Springer Berlin
  Heidelberg},\ \bibinfo {year} {2006})\BibitemShut {NoStop}%
\bibitem [{\citenamefont {Kasamanyan}(1981)}]{zatik1}%
  \BibitemOpen
  \bibfield  {author} {\bibinfo {author} {\bibfnamefont {Z.~A.}\ \bibnamefont
  {Kasamanyan}},\ }\bibfield  {title} {\enquote {\bibinfo {title} {Local
  density of states in a model variable-band semiconductor},}\ }\href {\doibase
  10.1007/BF00892950} {\bibfield  {journal} {\bibinfo  {journal} {Soviet
  Physics Journal}\ }\textbf {\bibinfo {volume} {24}},\ \bibinfo {pages}
  {525--530} (\bibinfo {year} {1981})}\BibitemShut {NoStop}%
\bibitem [{\citenamefont {Gasparian}\ \emph {et~al.}(2005)\citenamefont
  {Gasparian}, \citenamefont {Altshuler},\ and\ \citenamefont
  {Ortu\~no}}]{GAO}%
  \BibitemOpen
  \bibfield  {author} {\bibinfo {author} {\bibfnamefont {V.}~\bibnamefont
  {Gasparian}}, \bibinfo {author} {\bibfnamefont {B.}~\bibnamefont
  {Altshuler}}, \ and\ \bibinfo {author} {\bibfnamefont {M.}~\bibnamefont
  {Ortu\~no}},\ }\bibfield  {title} {\enquote {\bibinfo {title} {Charge pumping
  in one-dimensional kronig-penney models},}\ }\href {\doibase
  10.1103/PhysRevB.72.195309} {\bibfield  {journal} {\bibinfo  {journal} {Phys.
  Rev. B}\ }\textbf {\bibinfo {volume} {72}},\ \bibinfo {pages} {195309}
  (\bibinfo {year} {2005})}\BibitemShut {NoStop}%
\bibitem [{\citenamefont {Aronov}\ \emph {et~al.}(1991)\citenamefont {Aronov},
  \citenamefont {Gasparian},\ and\ \citenamefont {Gummich}}]{aronov}%
  \BibitemOpen
  \bibfield  {author} {\bibinfo {author} {\bibfnamefont {A~G}\ \bibnamefont
  {Aronov}}, \bibinfo {author} {\bibfnamefont {V~M}\ \bibnamefont {Gasparian}},
  \ and\ \bibinfo {author} {\bibfnamefont {Ute}\ \bibnamefont {Gummich}},\
  }\bibfield  {title} {\enquote {\bibinfo {title} {Transmission of waves
  through one-dimensional random layered systems},}\ }\href {\doibase
  10.1088/0953-8984/3/17/017} {\bibfield  {journal} {\bibinfo  {journal}
  {Journal of Physics: Condensed Matter}\ }\textbf {\bibinfo {volume} {3}},\
  \bibinfo {pages} {3023} (\bibinfo {year} {1991})}\BibitemShut {NoStop}%
\bibitem [{\citenamefont {Gasparian}\ \emph
  {et~al.}(1997{\natexlab{b}})\citenamefont {Gasparian}, \citenamefont
  {Gummich}, \citenamefont {Jódar}, \citenamefont {Ruiz},\ and\ \citenamefont
  {Ortuño}}]{est}%
  \BibitemOpen
  \bibfield  {author} {\bibinfo {author} {\bibfnamefont {V.}~\bibnamefont
  {Gasparian}}, \bibinfo {author} {\bibfnamefont {Ute}\ \bibnamefont
  {Gummich}}, \bibinfo {author} {\bibfnamefont {E.}~\bibnamefont {Jódar}},
  \bibinfo {author} {\bibfnamefont {J.}~\bibnamefont {Ruiz}}, \ and\ \bibinfo
  {author} {\bibfnamefont {M.}~\bibnamefont {Ortuño}},\ }\bibfield  {title}
  {\enquote {\bibinfo {title} {Tunneling and dwell time for one-dimensional
  generalized kronig-penney model},}\ }\href {\doibase
  https://doi.org/10.1016/S0921-4526(96)01036-8} {\bibfield  {journal}
  {\bibinfo  {journal} {Physica B: Condensed Matter}\ }\textbf {\bibinfo
  {volume} {233}},\ \bibinfo {pages} {72--77} (\bibinfo {year}
  {1997}{\natexlab{b}})}\BibitemShut {NoStop}%
\bibitem [{\citenamefont {Smith}\ and\ \citenamefont {Principi}(2019)}]{benj}%
  \BibitemOpen
  \bibfield  {author} {\bibinfo {author} {\bibfnamefont {Thomas~Benjamin}\
  \bibnamefont {Smith}}\ and\ \bibinfo {author} {\bibfnamefont {Alessandro}\
  \bibnamefont {Principi}},\ }\bibfield  {title} {\enquote {\bibinfo {title} {A
  bipartite kronig–penney model with dirac-delta potential scatterers},}\
  }\href {\doibase 10.1088/1361-648X/ab4d67} {\bibfield  {journal} {\bibinfo
  {journal} {Journal of Physics: Condensed Matter}\ }\textbf {\bibinfo {volume}
  {32}},\ \bibinfo {pages} {055502} (\bibinfo {year} {2019})}\BibitemShut
  {NoStop}%
\bibitem [{\citenamefont {Reshodko}\ \emph {et~al.}(2019)\citenamefont
  {Reshodko}, \citenamefont {Benseny}, \citenamefont {Romhányi},\ and\
  \citenamefont {Busch}}]{resh}%
  \BibitemOpen
  \bibfield  {author} {\bibinfo {author} {\bibfnamefont {Irina}\ \bibnamefont
  {Reshodko}}, \bibinfo {author} {\bibfnamefont {Albert}\ \bibnamefont
  {Benseny}}, \bibinfo {author} {\bibfnamefont {Judit}\ \bibnamefont
  {Romhányi}}, \ and\ \bibinfo {author} {\bibfnamefont {Thomas}\ \bibnamefont
  {Busch}},\ }\bibfield  {title} {\enquote {\bibinfo {title} {Topological
  states in the kronig–penney model with arbitrary scattering potentials},}\
  }\href {\doibase 10.1088/1367-2630/aaf9bf} {\bibfield  {journal} {\bibinfo
  {journal} {New Journal of Physics}\ }\textbf {\bibinfo {volume} {21}},\
  \bibinfo {pages} {013010} (\bibinfo {year} {2019})}\BibitemShut {NoStop}%
\bibitem [{\citenamefont {Belloni}\ and\ \citenamefont
  {Robinett}(2014)}]{many}%
  \BibitemOpen
  \bibfield  {author} {\bibinfo {author} {\bibfnamefont {M.}~\bibnamefont
  {Belloni}}\ and\ \bibinfo {author} {\bibfnamefont {R.W.}\ \bibnamefont
  {Robinett}},\ }\bibfield  {title} {\enquote {\bibinfo {title} {The infinite
  well and dirac delta function potentials as pedagogical, mathematical and
  physical models in quantum mechanics},}\ }\href {\doibase
  https://doi.org/10.1016/j.physrep.2014.02.005} {\bibfield  {journal}
  {\bibinfo  {journal} {Physics Reports}\ }\textbf {\bibinfo {volume} {540}},\
  \bibinfo {pages} {25--122} (\bibinfo {year} {2014})},\ \bibinfo {note} {the
  infinite well and Dirac delta function potentials as pedagogical,
  mathematical and physical models in quantum mechanics}\BibitemShut {NoStop}%
\bibitem [{\citenamefont {Pedram}\ and\ \citenamefont {Vahabi}(2010)}]{iran}%
  \BibitemOpen
  \bibfield  {author} {\bibinfo {author} {\bibfnamefont {Pouria}\ \bibnamefont
  {Pedram}}\ and\ \bibinfo {author} {\bibfnamefont {M.}~\bibnamefont
  {Vahabi}},\ }\bibfield  {title} {\enquote {\bibinfo {title} {Exact solutions
  of a particle in a box with a delta function potential: The factorization
  method},}\ }\href {\doibase 10.1119/1.3373925} {\bibfield  {journal}
  {\bibinfo  {journal} {American Journal of Physics}\ }\textbf {\bibinfo
  {volume} {78}},\ \bibinfo {pages} {839--841} (\bibinfo {year} {2010})},\
  \Eprint
  {http://arxiv.org/abs/https://pubs.aip.org/aapt/ajp/article-pdf/78/8/839/13098733/839\_1\_online.pdf}
  {https://pubs.aip.org/aapt/ajp/article-pdf/78/8/839/13098733/839\_1\_online.pdf}
  \BibitemShut {NoStop}%
\bibitem [{\citenamefont {{De Lange}}\ and\ \citenamefont
  {Janssen}(1984)}]{kp1}%
  \BibitemOpen
  \bibfield  {author} {\bibinfo {author} {\bibfnamefont {C.}~\bibnamefont {{De
  Lange}}}\ and\ \bibinfo {author} {\bibfnamefont {T.}~\bibnamefont
  {Janssen}},\ }\bibfield  {title} {\enquote {\bibinfo {title} {Modulated
  kronig-penney model in superspace},}\ }\href {\doibase
  https://doi.org/10.1016/0378-4371(84)90123-7} {\bibfield  {journal} {\bibinfo
   {journal} {Physica A: Statistical Mechanics and its Applications}\ }\textbf
  {\bibinfo {volume} {127}},\ \bibinfo {pages} {125--140} (\bibinfo {year}
  {1984})}\BibitemShut {NoStop}%
\bibitem [{\citenamefont {Yuce}(2015)}]{report}%
  \BibitemOpen
  \bibfield  {author} {\bibinfo {author} {\bibfnamefont {C.}~\bibnamefont
  {Yuce}},\ }\bibfield  {title} {\enquote {\bibinfo {title} {Topological phase
  in a non-hermitian pt symmetric system},}\ }\href {\doibase
  https://doi.org/10.1016/j.physleta.2015.02.011} {\bibfield  {journal}
  {\bibinfo  {journal} {Physics Letters A}\ }\textbf {\bibinfo {volume}
  {379}},\ \bibinfo {pages} {1213--1218} (\bibinfo {year} {2015})}\BibitemShut
  {NoStop}%
\bibitem [{\citenamefont {Oksengendler}\ \emph {et~al.}(2017)\citenamefont
  {Oksengendler}, \citenamefont {Nikiforov},\ and\ \citenamefont
  {Maksimov}}]{tamm}%
  \BibitemOpen
  \bibfield  {author} {\bibinfo {author} {\bibfnamefont {B.~L.}\ \bibnamefont
  {Oksengendler}}, \bibinfo {author} {\bibfnamefont {V.~N.}\ \bibnamefont
  {Nikiforov}}, \ and\ \bibinfo {author} {\bibfnamefont {S.~E.}\ \bibnamefont
  {Maksimov}},\ }\bibfield  {title} {\enquote {\bibinfo {title} {Tamm states of
  fractal surfaces},}\ }\href {\doibase 10.1134/S1028335817060039} {\bibfield
  {journal} {\bibinfo  {journal} {Doklady Physics}\ }\textbf {\bibinfo {volume}
  {62}},\ \bibinfo {pages} {281--283} (\bibinfo {year} {2017})}\BibitemShut
  {NoStop}%
\bibitem [{\citenamefont {Ahmed}\ \emph {et~al.}(2016)\citenamefont {Ahmed},
  \citenamefont {Nathan},\ and\ \citenamefont {Ghosh}}]{ahmed}%
  \BibitemOpen
  \bibfield  {author} {\bibinfo {author} {\bibfnamefont {Zafar}\ \bibnamefont
  {Ahmed}}, \bibinfo {author} {\bibfnamefont {Joseph~Amal}\ \bibnamefont
  {Nathan}}, \ and\ \bibinfo {author} {\bibfnamefont {Dona}\ \bibnamefont
  {Ghosh}},\ }\bibfield  {title} {\enquote {\bibinfo {title} {Transparency of
  the complex pt-symmetric potentials for coherent injection},}\ }\href
  {\doibase https://doi.org/10.1016/j.physleta.2015.12.005} {\bibfield
  {journal} {\bibinfo  {journal} {Physics Letters A}\ }\textbf {\bibinfo
  {volume} {380}},\ \bibinfo {pages} {562--566} (\bibinfo {year}
  {2016})}\BibitemShut {NoStop}%
\bibitem [{\citenamefont {Ahmed}(2001)}]{ahm}%
  \BibitemOpen
  \bibfield  {author} {\bibinfo {author} {\bibfnamefont {Zafar}\ \bibnamefont
  {Ahmed}},\ }\bibfield  {title} {\enquote {\bibinfo {title} {Energy band
  structure due to a complex, periodic, pt-invariant potential},}\ }\href
  {\doibase https://doi.org/10.1016/S0375-9601(01)00426-1} {\bibfield
  {journal} {\bibinfo  {journal} {Physics Letters A}\ }\textbf {\bibinfo
  {volume} {286}},\ \bibinfo {pages} {231--235} (\bibinfo {year}
  {2001})}\BibitemShut {NoStop}%
\bibitem [{\citenamefont {Peyruchat}\ \emph {et~al.}(2024)\citenamefont
  {Peyruchat}, \citenamefont {Rodriguez}, \citenamefont {Smirr}, \citenamefont
  {Leone},\ and\ \citenamefont {Girit}}]{24}%
  \BibitemOpen
  \bibfield  {author} {\bibinfo {author} {\bibfnamefont {L.}~\bibnamefont
  {Peyruchat}}, \bibinfo {author} {\bibfnamefont {R.~H.}\ \bibnamefont
  {Rodriguez}}, \bibinfo {author} {\bibfnamefont {J.-L.}\ \bibnamefont
  {Smirr}}, \bibinfo {author} {\bibfnamefont {R.}~\bibnamefont {Leone}}, \ and\
  \bibinfo {author} {\bibfnamefont {\ifmmode \mbox{\c{C}}\else
  \c{C}\fi{}.~\"O.}\ \bibnamefont {Girit}},\ }\bibfield  {title} {\enquote
  {\bibinfo {title} {Spectral signatures of nontrivial topology in a
  superconducting circuit},}\ }\href {\doibase 10.1103/PhysRevX.14.041041}
  {\bibfield  {journal} {\bibinfo  {journal} {Phys. Rev. X}\ }\textbf {\bibinfo
  {volume} {14}},\ \bibinfo {pages} {041041} (\bibinfo {year}
  {2024})}\BibitemShut {NoStop}%
\bibitem [{\citenamefont {{Esaki}}\ \emph {et~al.}(2011)\citenamefont
  {{Esaki}}, \citenamefont {{Sato}}, \citenamefont {{Hasebe}},\ and\
  \citenamefont {{Kohmoto}}}]{esaki}%
  \BibitemOpen
  \bibfield  {author} {\bibinfo {author} {\bibfnamefont {Kenta}\ \bibnamefont
  {{Esaki}}}, \bibinfo {author} {\bibfnamefont {Masatoshi}\ \bibnamefont
  {{Sato}}}, \bibinfo {author} {\bibfnamefont {Kazuki}\ \bibnamefont
  {{Hasebe}}}, \ and\ \bibinfo {author} {\bibfnamefont {Mahito}\ \bibnamefont
  {{Kohmoto}}},\ }\bibfield  {title} {\enquote {\bibinfo {title} {{Edge states
  and topological phases in non-Hermitian systems}},}\ }\href {\doibase
  10.1103/PhysRevB.84.205128} {\bibfield  {journal} {\bibinfo  {journal}
  {\prb}\ }\textbf {\bibinfo {volume} {84}},\ \bibinfo {eid} {205128} (\bibinfo
  {year} {2011})},\ \Eprint {http://arxiv.org/abs/1107.2079} {arXiv:1107.2079
  [cond-mat.mes-hall]} \BibitemShut {NoStop}%
\bibitem [{\citenamefont {Chong}\ \emph {et~al.}(2011)\citenamefont {Chong},
  \citenamefont {Ge},\ and\ \citenamefont {Stone}}]{stonea}%
  \BibitemOpen
  \bibfield  {author} {\bibinfo {author} {\bibfnamefont {Y.~D.}\ \bibnamefont
  {Chong}}, \bibinfo {author} {\bibfnamefont {Li}~\bibnamefont {Ge}}, \ and\
  \bibinfo {author} {\bibfnamefont {A.~Douglas}\ \bibnamefont {Stone}},\
  }\bibfield  {title} {\enquote {\bibinfo {title}
  {$\mathcal{P}\mathcal{T}$-symmetry breaking and laser-absorber modes in
  optical scattering systems},}\ }\href {\doibase
  10.1103/PhysRevLett.106.093902} {\bibfield  {journal} {\bibinfo  {journal}
  {Phys. Rev. Lett.}\ }\textbf {\bibinfo {volume} {106}},\ \bibinfo {pages}
  {093902} (\bibinfo {year} {2011})}\BibitemShut {NoStop}%
\bibitem [{\citenamefont {Achilleos}\ \emph {et~al.}(2017)\citenamefont
  {Achilleos}, \citenamefont {Aur\'egan},\ and\ \citenamefont {Pagneux}}]{FS}%
  \BibitemOpen
  \bibfield  {author} {\bibinfo {author} {\bibfnamefont {V.}~\bibnamefont
  {Achilleos}}, \bibinfo {author} {\bibfnamefont {Y.}~\bibnamefont
  {Aur\'egan}}, \ and\ \bibinfo {author} {\bibfnamefont {V.}~\bibnamefont
  {Pagneux}},\ }\bibfield  {title} {\enquote {\bibinfo {title} {Scattering by
  finite periodic $\mathcal{P}\mathcal{T}$-symmetric structures},}\ }\href
  {\doibase 10.1103/PhysRevLett.119.243904} {\bibfield  {journal} {\bibinfo
  {journal} {Phys. Rev. Lett.}\ }\textbf {\bibinfo {volume} {119}},\ \bibinfo
  {pages} {243904} (\bibinfo {year} {2017})}\BibitemShut {NoStop}%
\bibitem [{\citenamefont {Burke}\ \emph
  {et~al.}(2020{\natexlab{b}})\citenamefont {Burke}, \citenamefont {Wiersig},\
  and\ \citenamefont {Haque}}]{japan}%
  \BibitemOpen
  \bibfield  {author} {\bibinfo {author} {\bibfnamefont {Phillip~C.}\
  \bibnamefont {Burke}}, \bibinfo {author} {\bibfnamefont {Jan}\ \bibnamefont
  {Wiersig}}, \ and\ \bibinfo {author} {\bibfnamefont {Masudul}\ \bibnamefont
  {Haque}},\ }\bibfield  {title} {\enquote {\bibinfo {title} {Non-hermitian
  scattering on a tight-binding lattice},}\ }\href {\doibase
  10.1103/PhysRevA.102.012212} {\bibfield  {journal} {\bibinfo  {journal}
  {Phys. Rev. A}\ }\textbf {\bibinfo {volume} {102}},\ \bibinfo {pages}
  {012212} (\bibinfo {year} {2020}{\natexlab{b}})}\BibitemShut {NoStop}%
\bibitem [{\citenamefont {Garmon}\ \emph {et~al.}(2015)\citenamefont {Garmon},
  \citenamefont {Gianfreda},\ and\ \citenamefont {Hatano}}]{japan1}%
  \BibitemOpen
  \bibfield  {author} {\bibinfo {author} {\bibfnamefont {Savannah}\
  \bibnamefont {Garmon}}, \bibinfo {author} {\bibfnamefont {Mariagiovanna}\
  \bibnamefont {Gianfreda}}, \ and\ \bibinfo {author} {\bibfnamefont
  {Naomichi}\ \bibnamefont {Hatano}},\ }\bibfield  {title} {\enquote {\bibinfo
  {title} {Bound states, scattering states, and resonant states in
  $\mathcal{PT}$-symmetric open quantum systems},}\ }\href {\doibase
  10.1103/PhysRevA.92.022125} {\bibfield  {journal} {\bibinfo  {journal} {Phys.
  Rev. A}\ }\textbf {\bibinfo {volume} {92}},\ \bibinfo {pages} {022125}
  (\bibinfo {year} {2015})}\BibitemShut {NoStop}%
\bibitem [{\citenamefont {Shobe}\ \emph {et~al.}(2021)\citenamefont {Shobe},
  \citenamefont {Kuramoto}, \citenamefont {Imura},\ and\ \citenamefont
  {Hatano}}]{japan2}%
  \BibitemOpen
  \bibfield  {author} {\bibinfo {author} {\bibfnamefont {Ken}\ \bibnamefont
  {Shobe}}, \bibinfo {author} {\bibfnamefont {Keiichi}\ \bibnamefont
  {Kuramoto}}, \bibinfo {author} {\bibfnamefont {Ken-Ichiro}\ \bibnamefont
  {Imura}}, \ and\ \bibinfo {author} {\bibfnamefont {Naomichi}\ \bibnamefont
  {Hatano}},\ }\bibfield  {title} {\enquote {\bibinfo {title} {Non-hermitian
  fabry-p\'erot resonances in a $pt$-symmetric system},}\ }\href {\doibase
  10.1103/PhysRevResearch.3.013223} {\bibfield  {journal} {\bibinfo  {journal}
  {Phys. Rev. Res.}\ }\textbf {\bibinfo {volume} {3}},\ \bibinfo {pages}
  {013223} (\bibinfo {year} {2021})}\BibitemShut {NoStop}%
\bibitem [{\citenamefont {Jin}\ and\ \citenamefont {Song}(2009)}]{song}%
  \BibitemOpen
  \bibfield  {author} {\bibinfo {author} {\bibfnamefont {L.}~\bibnamefont
  {Jin}}\ and\ \bibinfo {author} {\bibfnamefont {Z.}~\bibnamefont {Song}},\
  }\bibfield  {title} {\enquote {\bibinfo {title} {Solutions of
  $\mathcal{P}\mathcal{T}$-symmetric tight-binding chain and its equivalent
  hermitian counterpart},}\ }\href {\doibase 10.1103/PhysRevA.80.052107}
  {\bibfield  {journal} {\bibinfo  {journal} {Phys. Rev. A}\ }\textbf {\bibinfo
  {volume} {80}},\ \bibinfo {pages} {052107} (\bibinfo {year}
  {2009})}\BibitemShut {NoStop}%
\bibitem [{\citenamefont {Muga}\ \emph {et~al.}(2004)\citenamefont {Muga},
  \citenamefont {Palao}, \citenamefont {Navarro},\ and\ \citenamefont
  {Egusquiza}}]{muga}%
  \BibitemOpen
  \bibfield  {author} {\bibinfo {author} {\bibfnamefont {J.G.}\ \bibnamefont
  {Muga}}, \bibinfo {author} {\bibfnamefont {J.P.}\ \bibnamefont {Palao}},
  \bibinfo {author} {\bibfnamefont {B.}~\bibnamefont {Navarro}}, \ and\
  \bibinfo {author} {\bibfnamefont {I.L.}\ \bibnamefont {Egusquiza}},\
  }\bibfield  {title} {\enquote {\bibinfo {title} {Complex absorbing
  potentials},}\ }\href {\doibase
  https://doi.org/10.1016/j.physrep.2004.03.002} {\bibfield  {journal}
  {\bibinfo  {journal} {Physics Reports}\ }\textbf {\bibinfo {volume} {395}},\
  \bibinfo {pages} {357--426} (\bibinfo {year} {2004})}\BibitemShut {NoStop}%
\end{thebibliography}%

\end{document}